\documentclass[twocolumn]{aastex631}

\received{12 January 2023}
\revised{29 June 2023}
\accepted{2 July 2023}
\published{22 August 2023}

\submitjournal{ApJ}

\begin{document}

\correspondingauthor{Craig Pellegrino}
\email{cpellegrino@lco.global}

\author[0000-0002-7472-1279]{C. Pellegrino}
\affil{Las Cumbres Observatory, 6740 Cortona Drive, Suite 102, Goleta, CA 93117-5575, USA}
\affil{Department of Physics, University of California, Santa Barbara, CA 93106-9530, USA}

\author[0000-0002-1125-9187]{D. Hiramatsu}
\affil{Center for Astrophysics \textbar Harvard \& Smithsonian, 60 Garden Street, Cambridge, MA 02138-1516, USA}
\affil{The NSF AI Institute for Artificial Intelligence and Fundamental Interactions}

\author[0000-0001-7090-4898]{I. Arcavi}
\affil{The School of Physics and Astronomy, Tel Aviv University, Tel Aviv 6997801, Israel}
\affil{CIFAR Azrieli Global Scholars Program, CIFAR, Toronto, Canada}

\author[0000-0003-4253-656X]{D. A. Howell}
\affil{Las Cumbres Observatory, 6740 Cortona Drive, Suite 102, Goleta, CA 93117-5575, USA}
\affil{Department of Physics, University of California, Santa Barbara, CA 93106-9530, USA}

\author[0000-0002-4924-444X]{K.\ A. Bostroem}
\altaffiliation{LSST Catalyst Fellow}
\affil{Department of Astronomy, University of Washington, 3910 15th Avenue NE, Seattle, WA 98195-0002, USA}

\author[0000-0001-6272-5507]{P. J. Brown}
\affil{Department of Physics and Astronomy, Texas A\&M University, 4242 TAMU, College Station, TX 77843, USA}
\affil{George P. and Cynthia Woods Mitchell Institute for Fundamental Physics \& Astronomy, College Station, TX 77843, USA}

\author[0000-0003-0035-6659]{J. Burke}
\affil{Las Cumbres Observatory, 6740 Cortona Drive, Suite 102, Goleta, CA 93117-5575, USA}
\affil{Department of Physics, University of California, Santa Barbara, CA 93106-9530, USA}

\author[0000-0002-1381-9125]{N. Elias-Rosa}
\affil{INAF - Osservatorio Astronomico di Padova, Vicolo
dell’Osservatorio 5, I-35122 Padova, Italy}
\affil{Institute of Space Sciences (ICE, CSIC), Campus UAB, Carrer
de Can Magrans s/n, 08193 Barcelona, Spain}

\author{K. Itagaki}
\affil{Itagaki Astronomical Observatory, Yamagata, Yamagata
990-2492, Japan}

\author{H. Kaneda}
\affil{Kaneda Astronomical Observatory, Sapporo, Hokkaido 005-0862, Japan}

\author[0000-0001-5807-7893]{C. McCully}
\affil{Las Cumbres Observatory, 6740 Cortona Drive, Suite 102, Goleta, CA 93117-5575, USA}
\affil{Department of Physics, University of California, Santa Barbara, CA 93106-9530, USA}

\author[0000-0001-7132-0333]{M. Modjaz}
\affil{Department of Astronomy, University of Virginia, Charlottesville, VA 22904, USA}

\author[0000-0003-0209-9246]{E. Padilla Gonzalez}
\affil{Las Cumbres Observatory, 6740 Cortona Drive, Suite 102, Goleta, CA 93117-5575, USA}
\affil{Department of Physics, University of California, Santa Barbara, CA 93106-9530, USA}

\author[0000-0001-9227-8349]{T. A. Pritchard}
\affil{Department of Physics, New York University, New York, NY 10003, USA}

\author{N. Yesmin}
\affil{Department of Astronomy, University of Virginia, Charlottesville, VA 22904, USA}

\title{SN 2020bio: A Double-peaked, H-poor Type IIb Supernova with Evidence of Circumstellar Interaction}

\shortauthors{Pellegrino et al.}
\shorttitle{The Double-peaked, H-poor Type IIb SN 2020bio}

\begin{abstract}

We present photometric and spectroscopic observations of SN\,2020bio, a double-peaked Type IIb supernova (SN) discovered within a day of explosion, primarily obtained by Las Cumbres Observatory and Swift. SN\,2020bio displays a rapid and long-lasting initial decline throughout the first week of its light curve, similarly to other well-studied Type IIb SNe. This early-time emission is thought to originate from the cooling of the extended outer hydrogen-rich (H-rich) envelope of the progenitor star that is shock heated by the SN explosion. We compare SN\,2020bio to a sample of other double-peaked Type IIb SNe in order to investigate its progenitor properties. Analytical model fits to the early-time emission give progenitor radius ($\approx$ 100--1500 $R_\odot$) and H-rich envelope mass ($\approx$ 0.01--0.5 $M_\odot$) estimates that are consistent with other Type IIb SNe. However, SN\,2020bio displays several peculiarities, including: (1) weak H spectral features indicating a greater amount of mass loss than other Type IIb progenitors; (2) an underluminous secondary light-curve peak that implies a small amount of synthesized $^{56}$Ni ($M_{\text{Ni}}$ $\approx$ 0.02 $M_\odot$); and (3) low-luminosity nebular [\ion{O}{1}] and interaction-powered nebular features. These observations are more consistent with a lower-mass progenitor ($M_{\text{ZAMS}} \approx$ 12 $M_\odot$) that was stripped of most of its H-rich envelope before exploding. This study adds to the growing diversity in the observed properties of Type IIb SNe and their progenitors.

\end{abstract}


\section{Introduction} \label{sec:intro}

While the majority of stars with initial masses $\gtrsim$ 8 $M_\odot$ end their lives as H-rich core-collapse supernovae \citep[SNe; e.g.,][]{Janka2012}, some massive stars lose their outer H and even He envelopes and explode as stripped-envelope SNe \citep[SESNe; e.g.,][]{Filippenko1997,GalYam2017}. A small but growing number of SNe have been observed with spectra that show similarities to both these classes \citep{Smith2011}. Classified as Type IIb SNe (SNe IIb), their spectra have H features at early times that gradually give way to He features, indicating that their progenitors were partially stripped of their outer envelopes before exploding \citep{Woosley1994}. 

It is unclear what mechanisms are responsible for this mass loss. Common hypotheses include stellar winds, binary interaction, or late-stage stellar instabilities \citep[see, e.g.,][for a review]{Smith2014}. Recent studies have shown that mass loss is common during the late stages of massive star evolution, as inferred from early-time observations of core-collapse SNe \citep[e.g.,][]{Ofek2014,Bruch2021,Strotjohann2021}. A significant fraction of core-collapse SNe show signatures of pre-existing circumstellar material (CSM) in their early-time spectra, obtained days after their estimated explosion epochs. This CSM is the material shed by the progenitor star in the months to years before core collapse. As the SN shock breaks out of the expanding ejecta, the resulting X-ray and ultraviolet (UV) flash may ionize the surrounding CSM, producing narrow spectral features as the CSM cools and recombines \citep[e.g.,][]{Fassia2001,Yaron2017}. Interaction between the SN ejecta and CSM can also influence the early-time light-curve evolution \citep{Morozova2018}.

Some SNe IIb are observed to have double-peaked light curves, with rapidly fading luminosities during the first several days after explosion before the radioactive decay of $^{56}$Ni synthesized during the explosion causes a rebrightening that lasts for several weeks. The early-time emission is thought to be the cooling of the extended envelope of the progenitor star that is heated by the SN shock \citep{Soderberg2012}. This shock-cooling emission (SCE) has only been extensively observed in a handful of cases, including SN\,1993J \citep[e.g.,][]{Richmond1994,Woosley1994}, SN\,2011dh \citep[e.g.,][]{Arcavi2011,Ergon2014}, SN\,2013df \citep[e.g.,][]{MG2014,VD2014}, SN\,2016gkg \citep{Arcavi2017}, SN\,2017jgh \citep{Armstrong2021}, and ZTF18aalrxas \citep{Fremling2019}, among others. Most of these objects are nearby and have had follow-up observations scheduled hours after explosion, which proved crucial to observing the rapidly evolving SCE. These studies have found that SNe IIb are consistent with the explosions of stars with extended outer envelopes, with the duration of the SCE dependent on the extent of this envelope \citep{Soderberg2012}.

Numerical and analytical models of SCE can complement pre-explosion imaging in determining the progenitors of these objects. Several models have been successful in reproducing the observed early-time evolution across all wavelengths. \citet[][hereafter P15]{Piro2015} is one of the first to present a one-zone analytical description of the cooling of an extended low-mass envelope shock-heated by the explosion of a compact massive core. \citet[][hereafter P21]{Piro2021} extend this to a two-zone model in order to better capture the emission from the outermost material in extended envelopes. \citet[][hereafter SW17]{Sapir2017} calibrate earlier models by \citet{Rabinak2011}\textemdash that depend on the precise density structure of the outer material\textemdash to numerical simulations for several days after explosion. 

Comparing observed SCE to analytical and numerical models is one of the only ways of directly measuring the radii and stellar structure of core-collapse progenitors from SN observations. This has been done for a handful of SNe IIb as well as SNe of other subtypes, including stripped-envelope Type Ib SNe \citep[e.g.,][]{Modjaz2009,Yao2020}, short-plateau Type II SNe \citep{Hiramatsu2021}, and exotic Ca-rich transients \citep[e.g.,][]{WJG2020,WJG2022}. Analytical and numerical modeling of double-peaked SNe IIb generally yield large radii progenitors ($\approx$ 100--500 $R_\odot$) with low-mass ($\approx$ 10$^{-2}$--10$^{-1}$ $M_\odot$) extended envelopes \citep[][and references therein]{Piro2021}. These properties are usually in agreement with those of SNe IIb progenitors from pre-explosion Hubble Space Telescope images, which have revealed them to be supergiants \citep{Aldering1994,Maund2011,VD2014}. In some cases, however, the progenitor radii estimated from SCE modeling are in tension with those measured from direct imaging \citep[e.g.,][in the case of SN\,2016gkg;]{Arcavi2017,Tartaglia2017}. Potential binary companions to the progenitor, which have been observed or inferred in a handful of cases \citep[e.g.,][]{Maund2004,Benvenuto2013} can further complicate direct imaging estimates when the individual binary members are unresolvable. 

Here, we present photometric and spectroscopic observations of SN\,2020bio, an SN IIb showing remarkably strong early-time emission, obtained by Las Cumbres Observatory (LCO) through the Global Supernova Project (GSP). LCO extensively observed SN\,2020bio from hours to $\approx$\,160 days after explosion, providing a detailed look into the full evolution of a double-peaked SN IIb. In this work, we analyze its light-curve evolution and spectral features, and we fit analytic models to its full light-curve evolution to estimate the radius, mass, and structure of its progenitor star. We also compare its bolometric light curve and spectra to numerical models in order to infer its progenitor mass and the properties of its circumstellar environment.

This paper is organized as follows. In Section \ref{sec:methods} we describe the discovery and follow-up observations of SN\,2020bio. We present its full light-curve and spectral time series in Section \ref{sec:analysis} and we compare observations to analytical and numerical models in Section \ref{sec:modelsec}. Finally, in Section \ref{sec:discussion} we discuss the potential progenitor properties of SN\,2020bio given the presented evidence.

\section{Discovery and Data Description}\label{sec:methods}

\begin{figure*}[t!]
    \centering
    \includegraphics[width=0.9\textwidth]{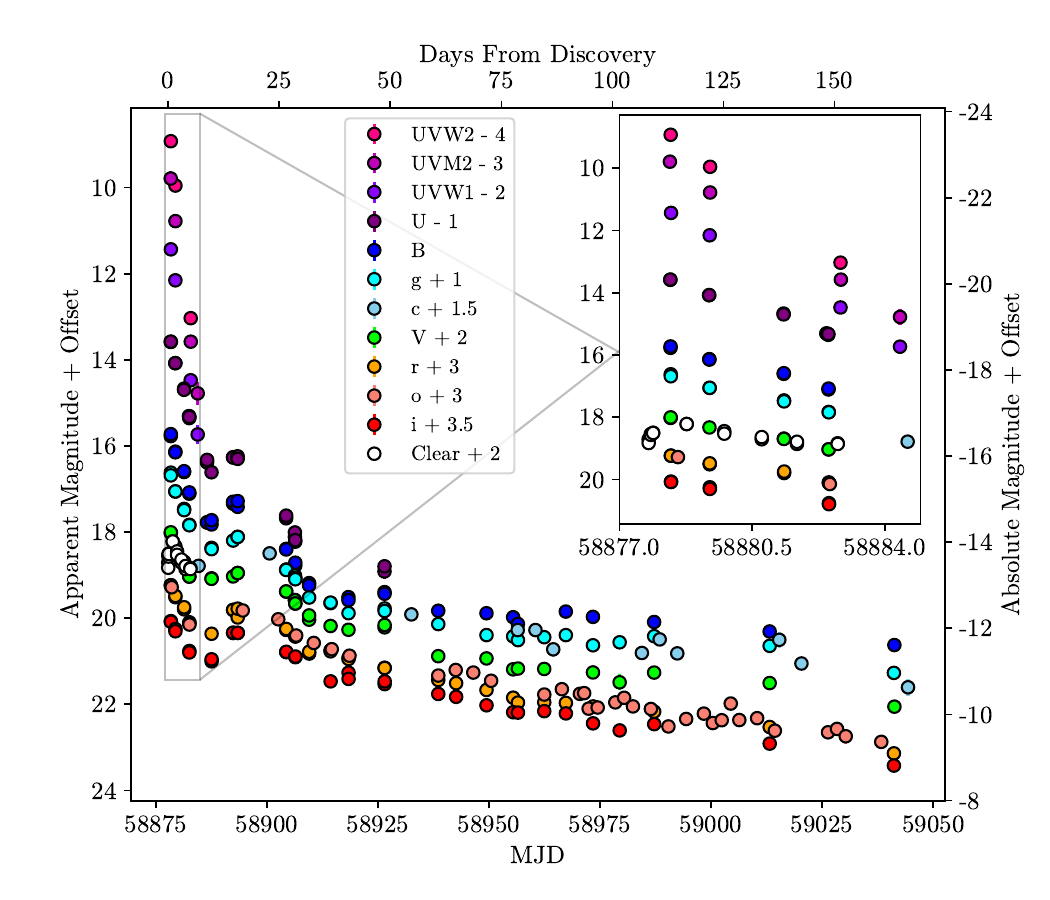}
    \caption{The full extinction-corrected light curves of SN\,2020bio. Photometry in different filters have been offset for clarity. Unfiltered photometry from the Itagaki Astronomical Observatory is included as clear points and calibrated to the \textit{V} band. The inset focuses on the rapidly evolving shock-cooling emission.}
    \label{fig:fulllc}
\end{figure*}

SN\,2020bio was discovered by Koichi Itagaki on UT 2020 January 29.77 at the Itagaki Astronomical Observatory at an unfiltered Vega magnitude of 16.7. Stacking images of the same field obtained by the Asteroid Terrestrial-impact Last Alert System (ATLAS) survey on the previous night yield a nondetection down to \textit{c}-band magnitude 20.6. Soon after discovery, rapid photometric and spectroscopic follow-up observations were requested by the GSP through the Las Cumbres global network of telescopes. The GSP also triggered its Swift Key Project (1518618: PI Howell) to obtain daily UV and optical photometry. A classification spectrum obtained on the 2.0m Liverpool Telescope on 2020 January 31.19\textemdash approximately 1.5 days after the first detection\textemdash shows a blue continuum superimposed with a narrow H$\alpha$ emission feature and a broad possible \ion{He}{1} $\lambda$ 5876\AA{} feature, consistent with a young core-collapse SN \citep{Srivastav2020}.

SN\,2020bio exploded at R.A. 13$^{\text{h}}$55$^{\text{m}}$37$^{\text{s}}$.69 and decl. +40\textdegree28$^\prime$39$^{\prime\prime}$.1 in the spiral galaxy NGC 5371 at redshift $z$ = 0.008533 \citep{Springob2005}. The distance to NGC 5371 is uncertain due to its low redshift. We adopt the mean of several distances measured using the method of \citet{Tully1977}, which gives $d$ = 29.9 $\pm$ 5.1 Mpc (values from the NASA Extragalactic Database\footnote{https://ned.ipac.caltech.edu/}). Using the \citet{Schlafly2011} dust map calibrations, we estimate a Galactic line-of-sight extinction to SN\,2020bio $E_{\text{MW}}(B-V)$ = 0.008 mag. Given the location of SN\,2020bio with respect to its host galaxy, we also estimate host extinction using the Na I D equivalent widths measured in a high-resolution spectrum of the SN. From the conversions presented in \citet{Poznanski2012}, we estimate $E_{\text{host}}(B-V)$\,=\,0.068 $\pm$ 0.038 mag for a total extinction $E(B-V)$ = 0.076 $\pm$ 0.038 mag. The photometry of SN\,2020bio presented throughout this work is corrected for this mean total extinction.

LCO photometric follow-up commenced less than a day after discovery. \textit{UBgVri}-band images were obtained by the Sinistro and Spectral cameras mounted on LCO 1.0 and 2.0 m telescopes, respectively, located at McDonald Observatory, Teide Observatory, and Haleakala Observatory. Data were reduced using \texttt{lcogtsnpipe} \citep{Valenti2016} which extracts point-spread function magnitudes after calculating zero points and color terms \citep{Stetson1987}. \textit{UBV}-band photometry was calibrated to Vega magnitudes using Landolt standard fields \citep{Landolt1992}, while \textit{gri}-band photometry was calibrated to AB magnitudes \citep{Smith2002} using Sloan Digital Sky Survey (SDSS) catalogs. As SN\,2020bio exploded coincident with its host galaxy, to remove host galaxy light we performed template subtraction using the HOTPANTS \citep{Becker2015} algorithm and template images obtained after the SN had faded. Unfiltered images were obtained with the Itagaki Astronomical Observatory (Okayama and Kochi, Japan) 0.35 m telescopes + KAF-1001E (CCD). Using our custom software, the photometry was extracted after host subtraction and calibrated to the \textit{V}-band magnitudes of 45 field stars from the Fourth US Naval Observatory CCD Astrograph Catalog \citep{Zacharias2013}.

We also obtained ATLAS \citep{Tonry2018,Smith2020} forced photometry from the forced photometry server \citep{Shingles2021}. Images obtained on the same night were averaged for higher signal-to-noise ratios. Magnitudes in the \textit{c} and \textit{o} bands were calibrated to AB magnitudes.

UV and optical photometry were obtained with the Ultraviolet and Optical Telescope \citep[UVOT;][]{Roming2005} on the Neil Gehrels Swift observatory \citep{Gehrels2004}. Swift data were reduced using a custom adaptation of the Swift Optical/Ultraviolet Supernova Archive \citep{Brown2014} pipeline with the most recent calibration files and the zero points of \citet{Breeveld2011}. Images from the final epoch, obtained after the SN had sufficiently faded, were used as templates to subtract the host galaxy light. All Swift photometry is calibrated to Vega magnitudes. The entire UV and optical data sets from LCO, Itagaki, and Swift UVOT are given in Table \ref{tab:phot} and shown in Figure \ref{fig:fulllc}.

\begin{deluxetable}{ccccc}[t]
\tablecaption{UV and Optical Photometry \label{tab:phot}}
\tablehead{
\colhead{JD} & \colhead{Filter} & \colhead{Magnitude} & \colhead{Uncertainty} & \colhead{Source}}
\startdata
2458878.27 & Clear & 16.77 & 0.15 & Itagaki \\
2458878.33 & Clear & 16.55 & 0.15 & Itagaki \\
2458878.39 & Clear & 16.51 & 0.15 & Itagaki \\
2458879.27 & Clear & 16.22 & 0.15 & Itagaki \\
2458880.26 & Clear & 16.49 & 0.15 & Itagaki \\
2458881.25 & Clear & 16.68 & 0.15 & Itagaki \\
2458882.18 & Clear & 16.82 & 0.15 & Itagaki \\
2458883.26 & Clear & 16.85 & 0.15 & Itagaki \\
2458878.85 & UVW2 & 13.56 & 0.04 & Swift \\
2458879.89 & UVW2 & 14.59 & 0.05 & Swift \\
\enddata
\tablenotetext{}{This table will be made available in its entirety in machine-readable format.}
\end{deluxetable}

LCO spectra were obtained by the FLOYDS spectrograph on the 2.0m Faulkes Telescope North at Haleakala Observatory. Spectra cover a wavelength range of 3500--10,000 \AA{} at a resolution $R$ $\approx$ 300--600. Data were reduced using the \texttt{floydsspec} pipeline,\footnote{https://github.com/svalenti/FLOYDS$\_$pipeline/} a custom pipeline, which performs cosmic ray removal, spectrum extraction, and wavelength and flux calibration. We also present one spectrum obtained by the B and C spectrograph on the 2.3m Bok Telescope at Steward Observatory, two spectra obtained by the Blue Channel Spectrograph on the 6.5m MMT at the Fred Lawrence Whipple Observatory, and one spectrum obtained by the Optical System for Imaging and low/intermediate-Resolution Integrated Spectroscopy (OSIRIS) spectrograph on the 10.4m Gran Telescopio Canarias. Details of all these spectra are presented in Table \ref{tab:speclog}.

\section{Photometric and Spectral Analysis}\label{sec:analysis}

\begin{figure*}
    \centering
    \includegraphics[width=0.67\textwidth]{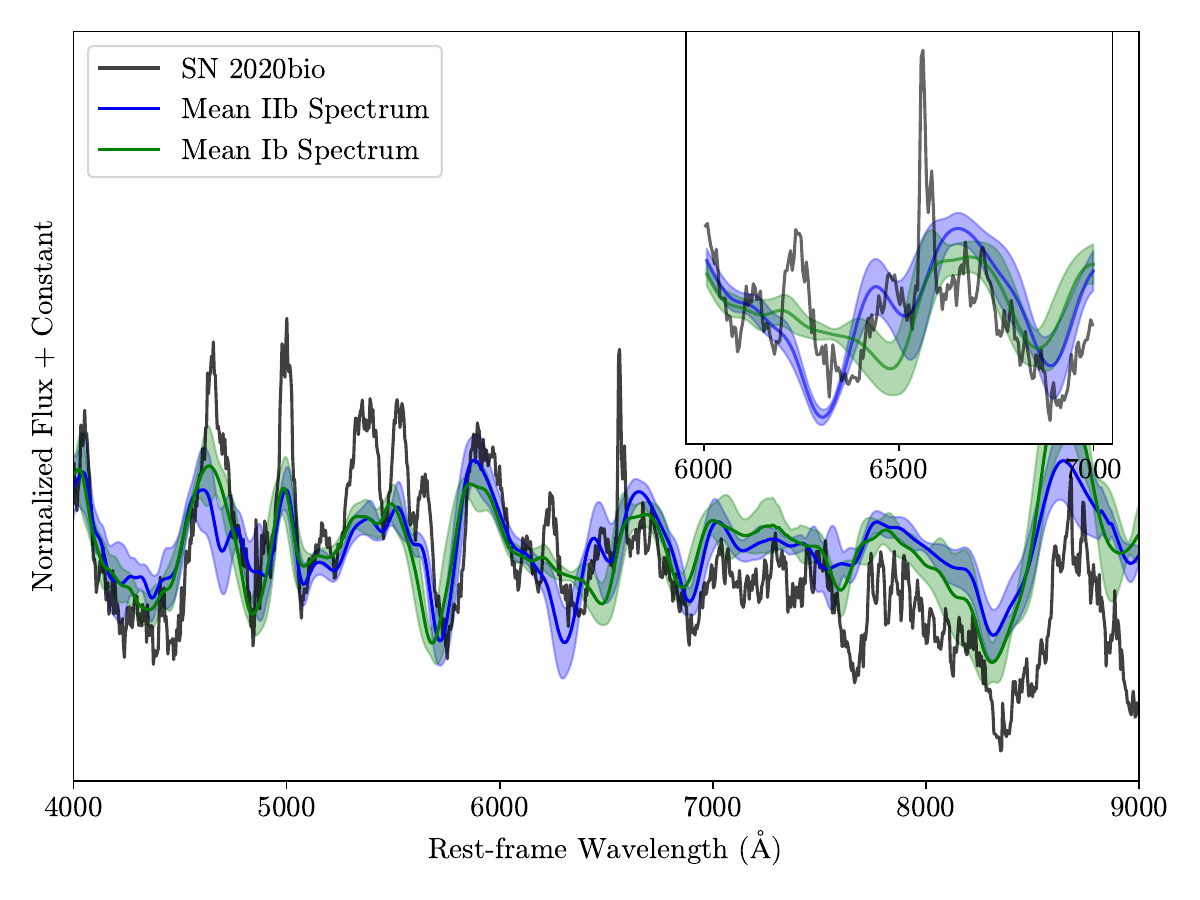}
    \caption{The spectrum of SN\,2020bio $\approx$26 days after explosion, compared with the mean Type Ib and Type IIb template spectra \citep{Liu2016} at the same phase. SN\,2020bio has spectral features in common with both these classes. The inset shows the region surrounding the H$\alpha$ line. SN\,2020bio has a weak and broad H$\alpha$ emission feature that is weaker than that in the mean SN IIb spectrum. However, this feature is much different than those seen in typical SNe Ib. This may indicate that the progenitor of SN\,2020bio was stripped almost entirely of its H-rich envelope.}
    \label{fig:classification}
\end{figure*}

\subsection{Spectroscopic Classification} \label{subsec:classification}
Given the unusual features of SN\,2020bio compared to SNe IIb in literature (see Sections \ref{sec:analysis} and \ref{sec:modelsec}), as well as the lack of a public classification, here we attempt to accurately classify SN\,2020bio. Analyzing the spectra of SN\,2020bio using SN classification software \citep{Howell2005,Blondin2007} gives matches to both SNe IIb and SNe Ib. At early times, more matches are found to SNe IIb, in particular the prototypical Type IIb SN\,1993J, than other classes (12 of 19 top \texttt{SNID} matches), while at later times, the matches are split more evenly. To explore the classification at this phase in greater detail, we compare the spectrum of SN\,2020bio 26 days after explosion with the mean Type Ib and Type IIb spectra from \citet{Liu2016} in Figure \ref{fig:classification}. As H$\alpha$ is one of the distinguishing features between SNe IIb and SNe Ib, we focus on this region of the spectra in the inset. SN\,2020bio lacks the strong, broad emission that is commonplace in most SNe IIb spectra. However, it also does not match the strong \ion{He}{1} $\lambda$ 6678 P Cygni feature that is seen in SNe Ib.

We consider two possibilities to explain the observed features in this region of the spectrum. First, it may consist mainly of broad but weak H$\alpha$ emission with a superimposed narrow host emission line. An absorption feature just blueward of that host line may be He $\lambda$ 6678 absorption, as seen in other ``flat-topped" H$\alpha$ features in SNe IIb. Assuming this absorption is from He, we measure an ejecta velocity of $\approx$ 7500 km s$^{-1}$. This velocity also corresponds to other absorption features seen in the spectrum corresponding to \ion{He}{1} $\lambda$ 5876 and \ion{He}{1} $\lambda$ 7065.

\begin{figure*} 
    \centering
    \includegraphics[width=0.65\textwidth]{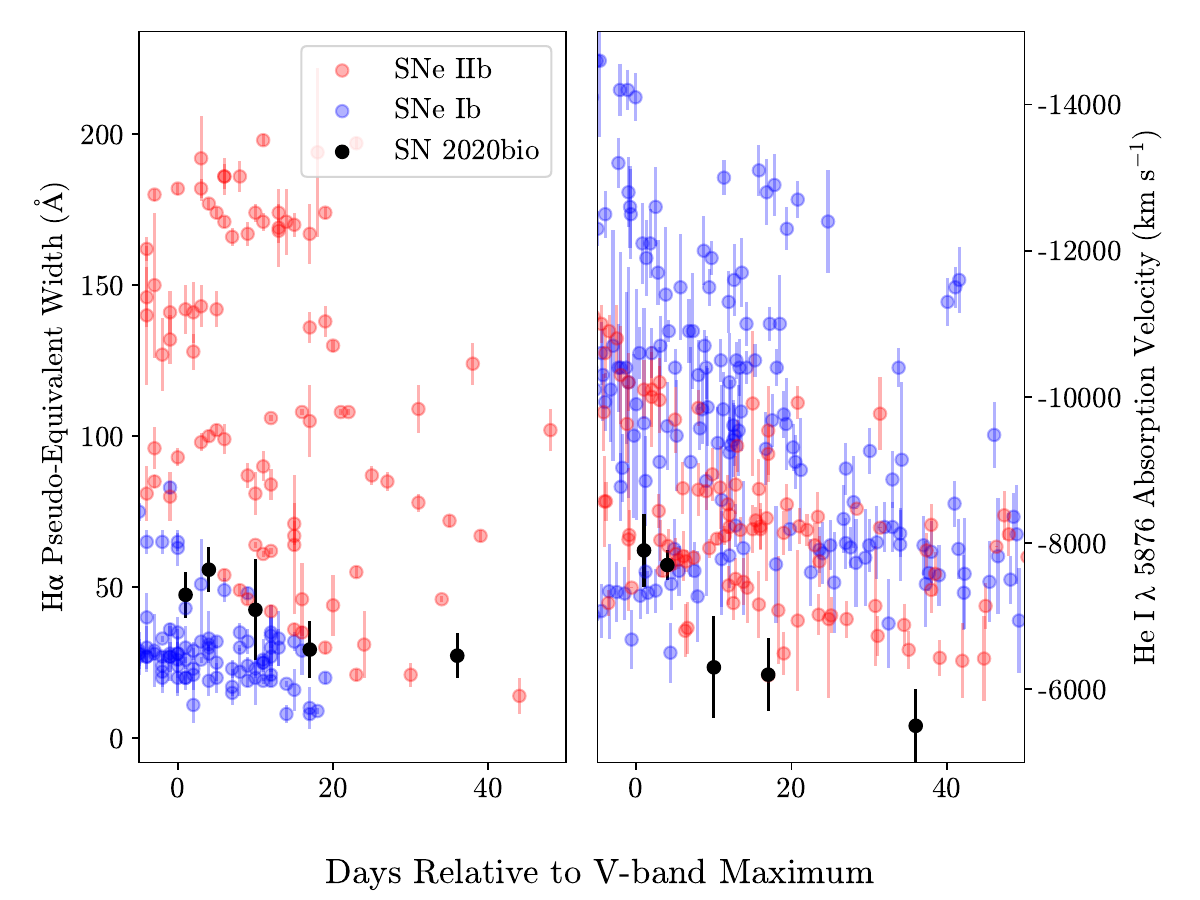}
    \caption{The evolution of the H$\alpha$ pseudo-equivalent width (left) and the \ion{He}{1} $\lambda$ 5876 absorption velocity (right) of SN\,2020bio (black) compared to SNe IIb (red) and Ib (blue) from \citet{Liu2016}. SN\,2020bio has values that are consistent with the relative extremes of both classes of objects but is more similar at later times to SNe IIb, in particular in its absorption velocities.}
    \label{fig:pewandvels}
\end{figure*}

The second possibility is that this region of the spectrum is dominated by a P Cygni feature of \ion{He}{1} $\lambda$ 6678. In this case, the He absorption from this P Cygni feature gives an ejecta velocity of $\approx$ 14,000 km s$^{-1}$. The narrow absorption just blueward of the narrow H$\alpha$ is more difficult to explain. One potential source is absorption by circumstellar H. The velocity of the absorption minimum relative to H$\alpha$ is $\approx$ 1000 km s$^{-1}$\textemdash faster than typical red supergiant or yellow supergiant winds \citep{Smith2014} but not unreasonable if the CSM is accelerated by interaction with the SN ejecta. Circumstellar interaction may also contribute to the narrow H$\alpha$ emission feature. While inspecting the 2D FLOYDS spectrum reveals residual H$\alpha$ contamination from the host galaxy, analyzing the full width at half maximum of this line over the first three weeks after explosion reveals a decreasing trend. Therefore, we cannot rule out circumstellar interaction as a contribution to the narrow H P Cygni feature.

Other diagnostics to differentiate between SNe IIb and Ib include the time evolution of the pseudo-equivalent width (pEW) of the H$\alpha$ absorption and the absorption velocity of the \ion{He}{1} $\lambda$ 5876 feature. \citet{Liu2016} show that both classes occupy relatively distinct regions of these parameter spaces. Here, we follow their methodology to measure both the H$\alpha$ pEW and the \ion{He}{1} absorption velocity in the spectra of SN\,2020bio beginning at the second \textit{V}-band peak. The results are plotted in Figure \ref{fig:pewandvels} along with the \citet{Liu2016} sample. Again, SN\,2020bio appears to be a transitional object, with similar pEW values to both SNe IIb and Ib. However, it also has much lower \ion{He}{1} absorption velocities than almost all the SNe Ib, in particular at late times. This is consistent with the conclusions of \citet{Liu2016}, who found that SNe IIb have systematically lower absorption velocities than SNe Ib throughout their evolution.

In summary, the classification of SN\,2020bio is difficult to determine with high confidence. The spectra reveal that this object is unique\textemdash emission from H-rich ejecta is very weak or nonexistent, and circumstellar interaction may be contributing to the peculiar spectral features. If there is weak but broad H emission, then the outer layers of the progenitor may have been almost entirely stripped of the H-rich material, placing SN\,2020bio in a transitional region between SNe IIb and SNe Ib. On the other hand, if the ejecta is H-free, then SN\,2020bio is a very rare example of an SN Ib with a double-peaked light curve. However, based on the emission centered around $\approx$6550 \AA{} that is broader than seen in typical SN Ib spectra, as well as the consistent ejecta velocity measurements of $\approx$ 7500 km s$^{-1}$, we favor the former interpretation. The close match to the SN IIb template, albeit with weak but broad H$\alpha$, allows us to classify this object as an SN of Type IIb.

\subsection{Light Curve and Color Evolution}

\begin{figure*}
    \centering
    \includegraphics[width=0.99\textwidth]{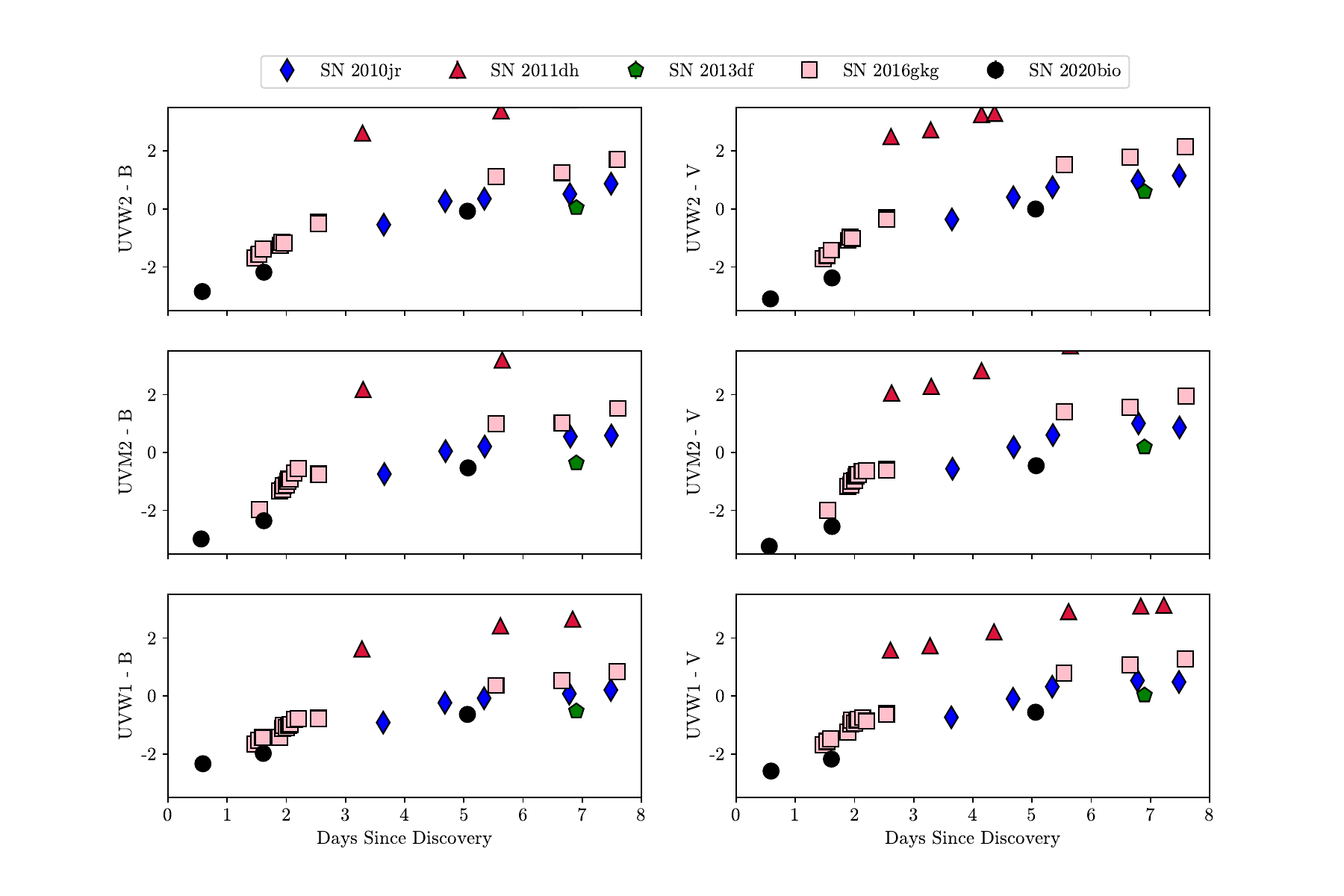}
    \caption{UV-optical colors of SN\,2020bio compared with those of other SNe IIb with early-time Swift observations. SN\,2020bio was bluer at earlier phases than the other SNe IIb. Data for these comparison SNe were obtained from the following sources: \citet{Arcavi2011} (SN\,2011dh), \citet{MG2014} (SN\,2013df), \citet{Arcavi2017} (SN\,2016gkg), \citet{Pritchard2014} and the Open Astronomy Catalog (SN 2010jr), and this work (SN\,2020bio).}
    \label{fig:uvcolors}
\end{figure*}
In Figure \ref{fig:fulllc} we show the full LCO, ATLAS, and Swift extinction-corrected light curve of SN\,2020bio, from detection to $\approx$ 160 days after explosion. The discovery and subsequent follow-up photometry from Itagaki are included as ``Clear" data points. The inset shows in greater detail the early-time evolution of the SCE, focusing on the first week after discovery. The most distinctive feature of the light curve is the luminous and rapidly declining SCE at early times. The peak SCE luminosity exceeds that of the secondary peak $\approx$ 15 days later, but SCE only dominates the light curve during the first several days. Over this time, the light curve falls by $\approx$ 4 mag in the first week, making this phase difficult to observe without rapid multiwavelength follow-up.

After $\approx$ 4 days from discovery, the slope of the light-curve decline changes as the luminosity from $^{56}$Ni decay begins to dominate the light curve. After about a week, the light curve rebrightens and reaches a secondary maximum $\approx$ 15 days after discovery. From this point, the emission settles onto the radioactive decay tail, powered by $^{56}$Co decay, for the remainder of the observations. The secondary peak and overall late-time light curve is relatively dim, peaking at $M$ $\approx$ -14 mag in the \textit{V}-band, hinting at a small amount of $^{56}$Ni synthesized in the explosion.

In Figure \ref{fig:uvcolors} we compare the early-time UV-optical colors of SN\,2020bio to those of other SNe IIb with observed SCE in the UV. The \textit{B-} and \textit{V}-band data for all the objects, with the exception of SN\,2010jr, consist of ground-based photometry in order to avoid uncertain subtractions and calibrations in Swift optical bands. All dates are given with respect to the time of discovery and corrected for extinction according to the published values for each object. SN\,2020bio has both the earliest observations relative to discovery and the bluest colors throughout its evolution compared to the other objects. While objects such as SN\,2010jr and SN\,2016gkg have more densely sampled light curves, their observations began later and their colors evolved redward faster compared to SN\,2020bio. Of the six colors plotted, SN\,2020bio is exceptionally blue in the \textit{UVM2-B} and \textit{UVM2-V} colors, in particular in the earliest epochs. This may be evidence for another luminosity contribution besides SCE, as we discuss in Section \ref{sec:discussion}.

\subsection{Spectral Comparison}

\begin{figure}[t!]
    \centering
    \includegraphics[width=0.47\textwidth]{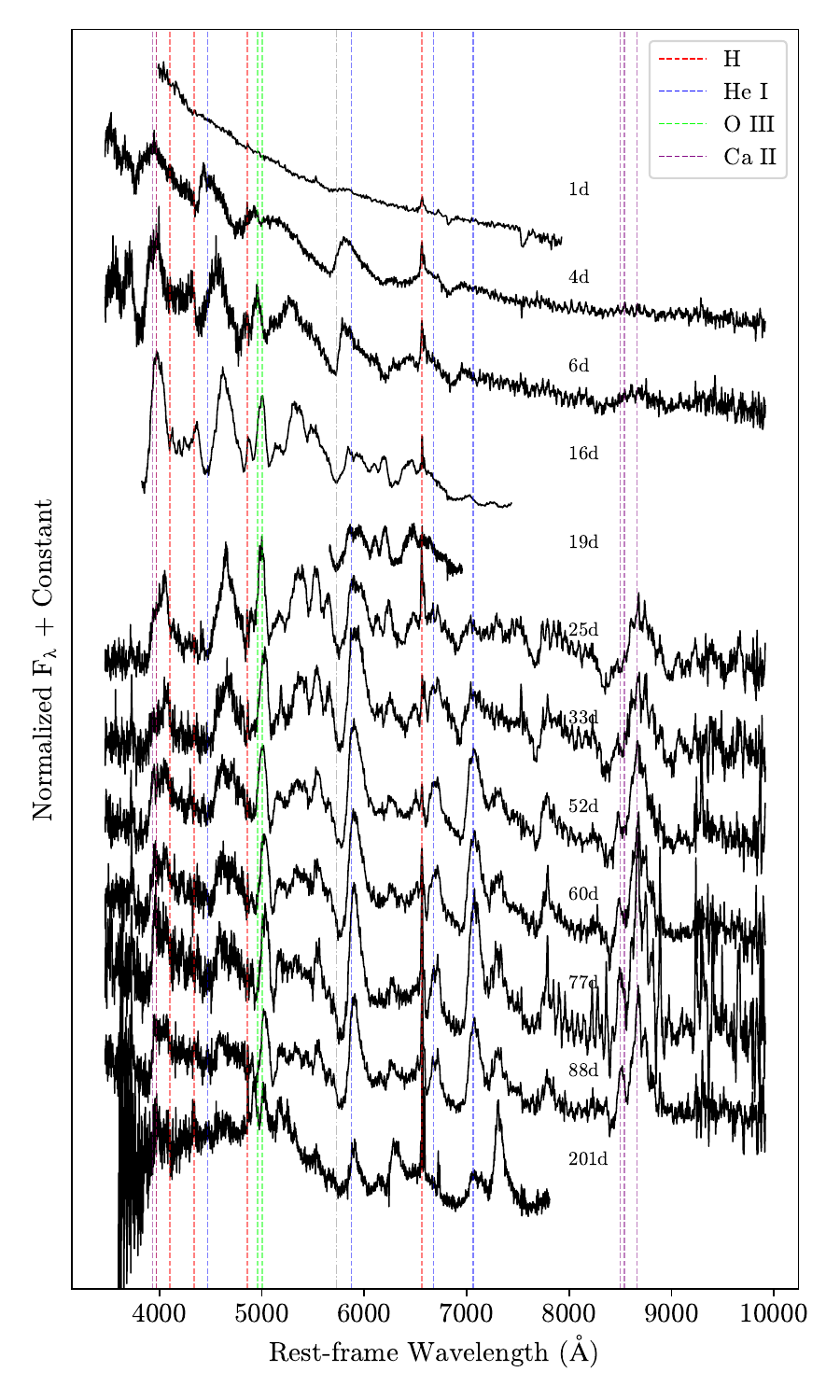}
    \caption{The full spectral time series of SN\,2020bio. Phases with respect to the detection epoch are given above each spectrum. Notable spectral features are identified with dashed lines. The gray dashed-dotted line shows a \ion{He}{1} $\lambda$ 5876 absorption velocity of 7500 km s$^{-1}$ for reference. The first spectrum is the publicly available classification spectrum retrieved from the Transient Name Server.}
    \label{fig:allspec}
\end{figure}

\begin{deluxetable*}{crlcc}[]
\tablecaption{Log of Spectroscopic Observations \label{tab:speclog}}
\tablehead{
\colhead{Date of Observation} & \colhead{Days since Discovery} & \colhead{Facility/Instrument} & \colhead{Exposure Time (s)} & 
\colhead{Wavelength Range (\AA{})}}
\startdata
 2020-01-31 04:27:31 & 1 & LT/SPRAT & 1200 & 4000--7925 \\
 2020-02-03 14:32:18 & 4 & LCO/FLOYDS-N & 1800 & 3500--10,000 \\
 2020-02-05 12:19:05 & 6 & LCO/FLOYDS-N & 1800 & 3500--10,000 \\
 2020-02-15 09:35:59 & 16 & Bok/B$\&$C & 600 & 3850--7500 \\
 2020-02-18 12:32:26 & 19 & MMT/Blue Channel & 300 & 5700--7000 \\
 2020-02-24 13:00:37 & 25 & LCO/FLOYDS-N & 1800 & 3500--10,000 \\
 2020-03-03 10:49:44 & 33 & LCO/FLOYDS-N & 2700 & 3500--10,000 \\
 2020-03-22 14:22:56 & 52 & LCO/FLOYDS-N & 3600 & 3500--10,000 \\
 2020-03-30 14:20:34 & 60 & LCO/FLOYDS-N & 3600 & 3500--10,000 \\
 2020-04-16 11:12:12 & 77 & LCO/FLOYDS-N & 3600 & 3500--10,000 \\
 2020-04-27 12:09:24 & 88 & LCO/FLOYDS-N & 3600 & 3500--10,000 \\
 2020-08-18 22:02:01 & 201 & GTC/OSIRIS & 1500 & 3600--7808 \\
\enddata
\tablecomments{All spectra will be made publicly available on WiseRep \citep{Yaron2012}.}
\end{deluxetable*}

Spectral coverage of SN\,2020bio began fewer than 2 days after the first detection\textemdash approximately 3 days since the estimated explosion time (Section \ref{subsec:models})\textemdash and continued for 201 days. We plot the full spectral series in Figure \ref{fig:allspec}. The earliest spectrum of SN\,2020bio, reported to the Transient Name Server \citep{Srivastav2020}, shows a hot blue continuum superimposed with weak emission features. We identify a potential weak, broad feature of \ion{He}{1} $\lambda$5876 \AA{} blueshifted by $\approx$ 11,000 km s$^{-1}$. We also note potential narrow lines of H$\alpha$ and H$\beta$; however, these features are consistent with host galaxy contamination at the resolution of the spectrum.

\begin{figure}
    \centering
    \includegraphics[width=0.47\textwidth]{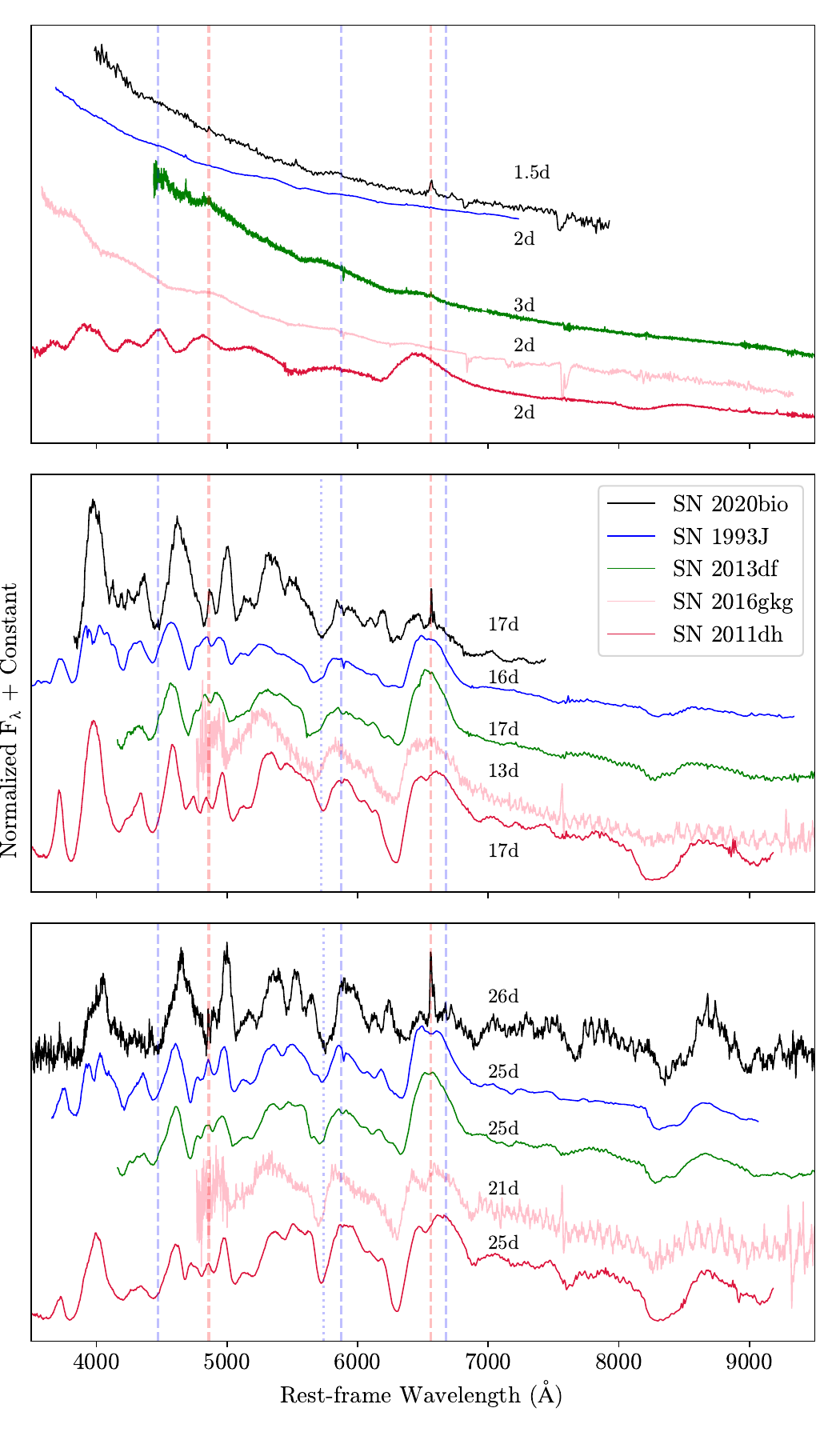}
    \caption{Spectra of SN\,2020bio compared with spectra of other SNe IIb at similar phases. Phases with respect to the estimated explosion time are given above each spectrum, and notable spectral features are identified with red (H) and blue (He) vertical lines at their rest-frame wavelengths. The dotted line shows the \ion{He}{1} $\lambda$ 5876 absorption velocity of SN\,2020bio. The spectra of SN\,2016gkg are unpublished spectra obtained by LCO, while the other comparison spectra were retrieved from WiseRep \citep{Yaron2012}.}
    \label{fig:comparespec}
\end{figure}

After about a week after the explosion, absorption features begin to develop in the spectra. We identify lines of He, O, and Ca. The absorption feature blueward of the rest-frame H$\alpha$ line matches \ion{He}{1} $\lambda$ 6678\AA{} absorption blueshifted by $\approx$ 7500 km s$^{-1}$, which is commonly noted to cause ``flat-topped" H$\alpha$ emission profiles in other SNe IIb \citep[e.g.,][]{Filippenko1993}. In general, the absorption features in the SN\,2020bio spectra are shallower than those of the other SNe IIb, in particular SN\,2011dh. Interaction with CSM can produce absorption features that are weaker and shallower than expected, which has been noted in the spectra of SN\,1993J and SN\,2013df \citep{Fremling2019}.

To further investigate the differences between SN\,2020bio and other SNe IIb, in Figure \ref{fig:comparespec} we plot comparison spectra just after (top), two weeks after (middle), and three weeks after (bottom) explosion. Among this sample, many of the other SNe IIb show broadened, high-velocity H and He features at early times. However, the spectrum of SN\,2020bio at this same phase shows only a blue continuum with possible weak \ion{He}{1} emission. This difference suggests that the photosphere has not yet receded within the outermost ejecta. One explanation for this is if the SN ejecta is surrounded by low-density CSM. At this phase, any narrow lines caused by photoionization or collisional excitation may have vanished, but the photosphere could still lie within this shock-heated material, obscuring the ejecta features within. By the time of our next spectrum (4 days after discovery) the photosphere has receded into the SN ejecta, revealing broad SN features. Narrow lines also persist in the spectra of SN\,2020bio; however, these are at least partially due to galaxy contamination, as \texttt{floydsspec} does not remove host galaxy contamination during the reduction.

Differences persist weeks after the estimated explosion times. While the other SNe IIb have developed broad H$\alpha$ and H$\beta$ emission features, these same lines are weaker in SN\,2020bio. This could be partly caused by \ion{He}{1} $\lambda$ 6678\AA{} absorption, which has an absorption trough coincident with the H$\alpha$ flux when blueshifted by $\approx$ 7500 km s$^{-1}$. Another possibility, as mentioned in Section \ref{subsec:classification}, is that the H emission from SN\,2020bio is inherently weaker than in other SNe IIb, which may be the case if the progenitor lost more of its outer H-rich envelope than the progenitors of the other SNe IIb did. Weak H emission, along with potential CSM, point to a scenario in which the progenitor of SN\,2020bio underwent enhanced mass loss, shedding almost all of its outer H layer before exploding. If this is the case, such a progenitor scenario for SN\,2020bio is unique among other well-studied SNe IIb.

\section{Light-curve Modeling and Progenitor Inference}\label{sec:modelsec}

\begin{figure*}
    \centering
    \includegraphics[width=0.49\textwidth]{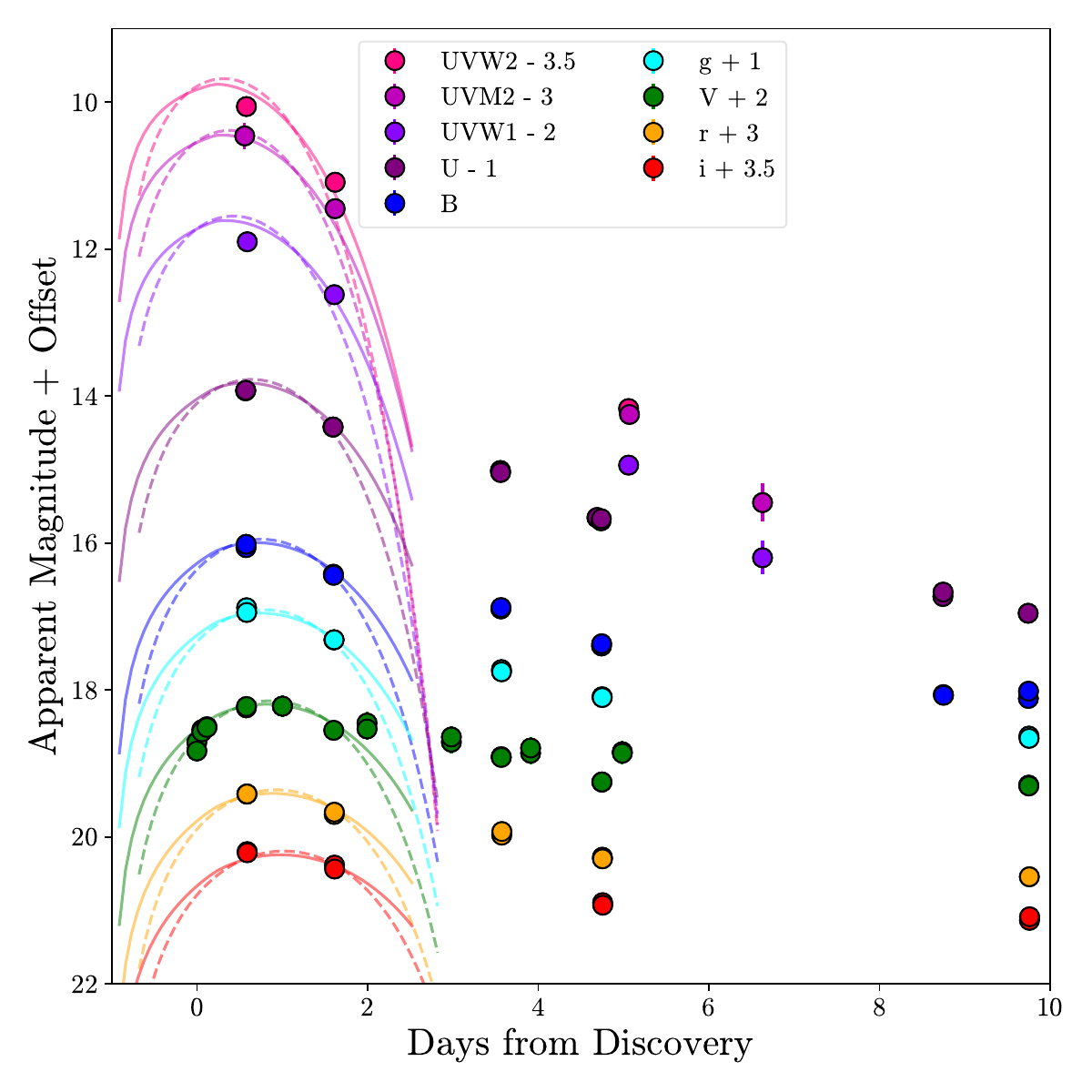}
    \includegraphics[width=0.49\textwidth]{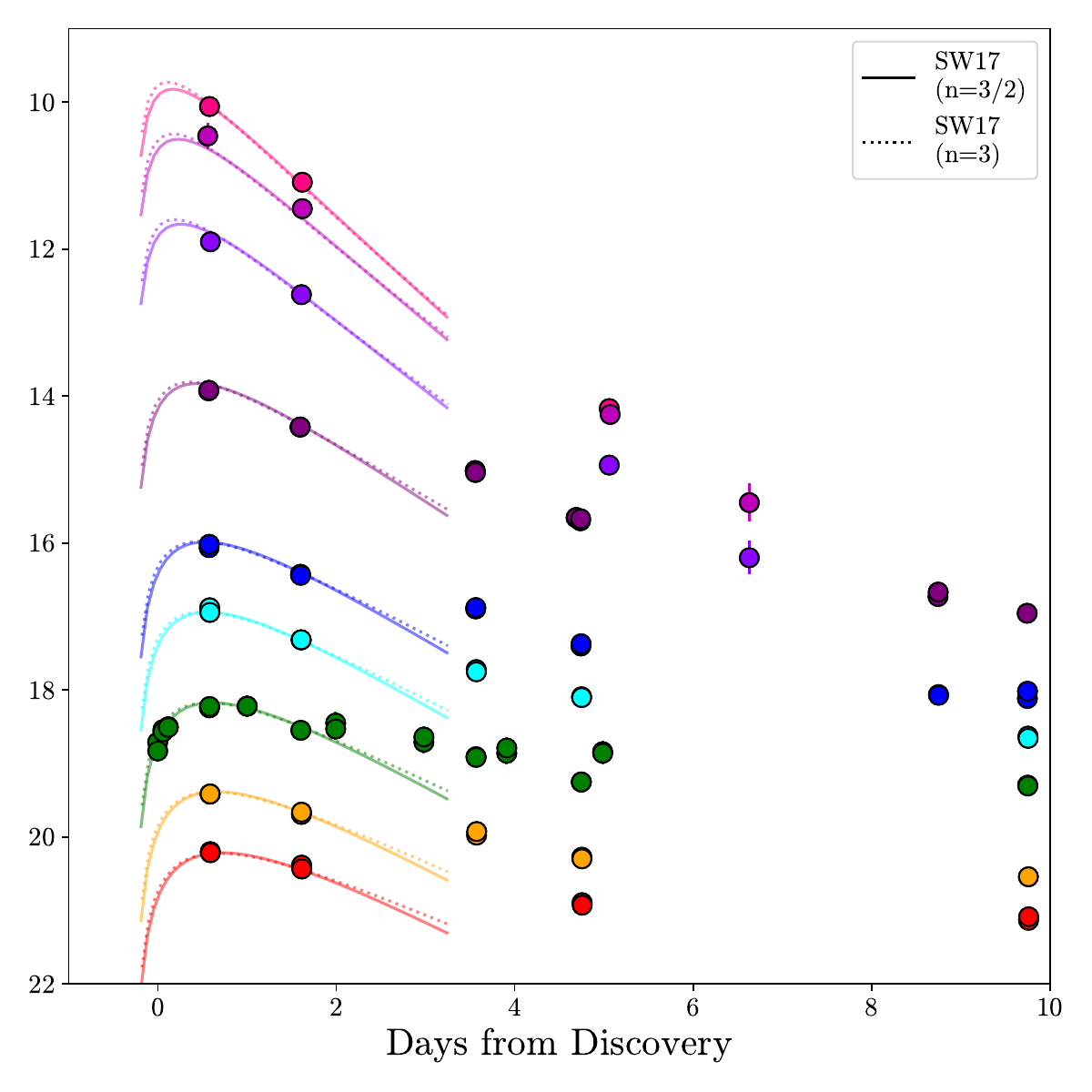}
    \caption{Shock-cooling fits to the early-time photometry of SN\,2020bio using the models of (left) P15 and P21 and that of (right) SW17, assuming a constant optical opacity appropriate for solar-composition material. Photometry in each band has been offset for clarity. Itagaki discovery photometry has been included in the \textit{V}-band fits.}
    \label{fig:scefits}
\end{figure*}

\begin{deluxetable*}{lccccc}[t!]
\tablecaption{SCE Model Parameters\label{tab:modelparams}}
\tablehead{
\colhead{Model} & \colhead{$R_{\text{env}}$ ($R_\odot$)} & \colhead{$M_{\text{env}}$ (10$^{-2}$ $M_\odot$)} & \colhead{$v^a$ (10$^{4}$ km s$^{-1}$)} & \colhead{$t_0$ (days)} & \colhead{$\chi^2$ / d.o.f.}}
\startdata
P15 & 510$_{-30}^{+30}$ & 1.14$_{-0.02}^{+0.02}$ & 1.67$_{-0.01}^{+0.02}$ & 0.67$_{-0.02}^{+0.02}$ & 21.6\\
P21 & 1700$_{-95}^{+85}$ & 1.60$_{-0.02}^{+0.03}$ & 1.36$_{-0.02}^{+0.01}$ &  0.98$_{-0.01}^{+0.01}$ & 21.1 \\
SW17 ($n$=3/2) & 160$_{-10}^{+12}$ & 47.12$_{-0.92}^{+0.96}$ & 1.69$_{-0.04}^{+0.04}$ & 0.26$_{-0.04}^{+0.04}$ & 8.7\\
SW17 ($n$=3) & 220$_{-15}^{+19}$ & 322.60$_{-6.20}^{+6.10}$ & 1.60$_{-0.04}^{+0.04}$ & 0.25$_{-0.04}^{+0.04}$ & 8.7\\
\enddata
\tablenotetext{a}{The characteristic velocity for P15 and P21 and the shock velocity for SW17.}
\end{deluxetable*}

\subsection{Shock-cooling Model Descriptions}

A variety of analytical and numerical models of SCE have been developed in recent years. Here, we consider three analytical models that are commonly used to fit the early-time emission of core-collapse SNe. The P15 model extends the formalism of \citet{Nakar2014} to reproduce the full shock-cooling peak. It assumes a lower-mass extended envelope without assuming its specific density structure. On the other hand, SW17 calibrates to the numerical models of \citet{Rabinak2011} and assumes specific polytropic indices for the extended envelope. The methodology used to fit these models to the data and derive resulting blackbody properties are presented in \citet{Arcavi2017}.

More recently, \citet{Piro2021} developed another analytical model to better reproduce the early SCE observed in a variety of transients \citep[e.g.,][]{Arcavi2017,Yao2020}. They assume a two-zone extended envelope in homologous expansion and calculate the emission from this shocked material. This method begins by assuming extended material in homologous expansion separated into two regions\textemdash an outer density profile described by $\rho$ $\propto$ $r^{-n}$, where n $\approx$ 10, and an inner region with $\rho$ $\propto$ $r^{-\delta}$, where $\delta$ $\approx$ 1.1. Assuming a transitional velocity $v_t$ between the inner and outer regions of the extended material, the time for the diffusion front to reach this transition is given by \begin{equation}
    t_d = \Bigl( \frac{3\kappa K M_e}{(n-1)v_t c} \Bigr)^{1/2}
\end{equation}
where $K$ = $\frac{(n-3)(3-\delta)}{4\pi(n-\delta)}$, $\kappa$ is the optical opacity, and $M_e$ is the mass of the extended material. The luminosity from the cooling of the extended material is then defined piecewise for times before and after this diffusion time: 

\begin{equation}
    L(t) \approx \frac{\pi(n-1)}{3(n-5)}\frac{cR_ev_t^2}{\kappa} \Bigl(\frac{t_d}{t}\Bigr)^{4/(n-2)}, t \leq t_d
\end{equation}
and 
\begin{equation}
    L(t) \approx \frac{\pi(n-1)}{3(n-5)}\frac{cR_ev_t^2}{\kappa}\text{exp}\Bigl[-\frac{1}{2}\Bigl(\frac{t^2}{t_d^2} - 1\Bigr)\Bigr] , t \geq t_d
\end{equation}

To fit the photometry in each band, we assume that the material radiates as a blackbody at some photospheric radius $r_{ph}$. The photosphere reaches the transition between the two regions at a time
\begin{equation}
    t_{ph} = \Bigl[\frac{3\kappa K M_e}{2(n-1)v_t^2}\Bigr]^{1/2}
\end{equation}
and the time evolution of the photospheric radius is given relative to this characteristic time:
\begin{equation}
    r_{ph}(t) = \Bigl(\frac{t_{ph}}{t}\Bigr)^{2/(n-1)}v_t t, t\leq t_{ph}
\end{equation}
and
\begin{equation}
    r_{ph}(t) = \Bigl[\frac{\delta - 1}{n - 1} \Bigl(\frac{t^2}{t_{ph}^2} - 1\Bigr) + 1 \Bigr]^{-1/(\delta - 1)}v_t t, t \geq t_{ph}
\end{equation}

In addition, we attempt to fit the analytical models of \citet{Shussman2016}, which are calibrated to numerical simulations from shock breakout to recombination. However, these model fits are unable to reproduce the rapidly declining shock-cooling emission in all filters during the week after explosion. It is possible this shortcoming is due to an unphysical application of the model\textemdash which is calibrated to numerical simulations of red supergiants\textemdash to the early light curve of SN\,2020bio, which likely had a different progenitor structure. Detailed comparisons between numerical models of SNe IIb and the \citet{Shussman2016} models are beyond the scope of this work.

\subsection{Best-fit Analytic Models}\label{subsec:models}

We fit each model to the early-time photometry of SN\,2020bio. For the SW17 model, we consider two polytropic indices ($n=3/2$ and $n=3$), appropriate for convective and radiative envelopes, respectively. Only data taken up to 3.5 days after discovery are fit, as this is the time when SCE dominates the luminosity over radioactive decay (see Section \ref{subsec:nilc} for a quantitative treatment of the $^{56}$Ni light curve). Additionally, we ensure that the phases we fit fall within the validity range of each model. In each case, we fit for the progenitor extended envelope radius, $R_{\text{env}}$, the envelope mass, $M_{\text{env}}$, either the characteristic velocity or the shock velocity $v$ of the outer material, and the offset time since explosion $t_{0}$. We use the \texttt{emcee} package \citep{ForemanMackey2013} to perform Markov Chain Monte Carlo fitting of each model, initializing 100 walkers with 1000 burn-in steps and running for an additional 1000 steps after burn-in. For each step, the total luminosity is computed using the analytical model formalism, and the luminosity within each filter is compared to the observed photometry assuming a blackbody spectral energy distribution (SED). We fit each model assuming an optical opacity $\kappa$ = 0.34 cm$^{2}$ g$^{-1}$, consistent with solar-composition material.

The best-fit models to the multiband SCE light curves are shown in Figure \ref{fig:scefits}, and best-fit parameters are given in Table \ref{tab:modelparams} with corner plots shown in Appendix \ref{sec:corners}. The Itagaki discovery data that capture the rise are calibrated to the \textit{V} band. We find that all the models fit the early-time data well, reproducing the rapid rise, luminous peak, and subsequent decline in all filters. Quantitatively, the SW17 model for convective envelopes ($n=3/2$) has the lowest reduced $\chi^2$ value, indicating the model most closely matches the observations. On the other hand, the best-fit envelope mass for the SW17 model with a radiative ($n=3$) envelope is larger than the total ejecta mass, estimated in Section \ref{subsec:nilc}. Therefore, we do not consider this model representative of the progenitor of SN\,2020bio.

Based on the unusual properties of SN\,2020bio compared to other SNe IIb, including its weak H spectral features and faint secondary light-curve peak, we test whether a lower-opacity envelope better reproduces the observed SCE. This could be the case if the progenitor star was almost completely stripped of its outer H-rich envelope. We perform the same fitting routine but fix the opacity $\kappa$\,=\,0.20 cm$^{2}$ g$^{-1}$ for H-poor material. The results are shown in Figure \ref{fig:hpoorfits}, with best-fit parameter values given in Table \ref{tab:hpoormodelparams}. We find no differences in goodness of fits for each model between the two chosen opacities\textemdash both the H-rich and H-poor envelopes produce similarly good fits. However, there are differences in the fitted parameters between the best-fit models. In the H-rich case, the envelope radii and masses from the best-fit SW17 model are consistent with those estimated for other SNe IIb (i.e. radii of $\approx$ 1$\times$10$^{13}$ cm and masses of 10$^{-3}$--10$^{-2}$ $M_\odot$). In the H-poor case, however, the radii are smaller ($\approx$ 100 $R_\odot$) and the envelope masses are larger ($\approx$ 0.5 $M_\odot$). These values are more consistent with those estimated for Type Ib and Ca-rich transients with observed SCE \citep[e.g.,][]{Yao2020,WJG2022}. 

\subsection{Bolometric Luminosities and Numerical Modeling}\label{subsec:nilc}

\begin{figure}
    \centering
    \includegraphics[width=0.48\textwidth]{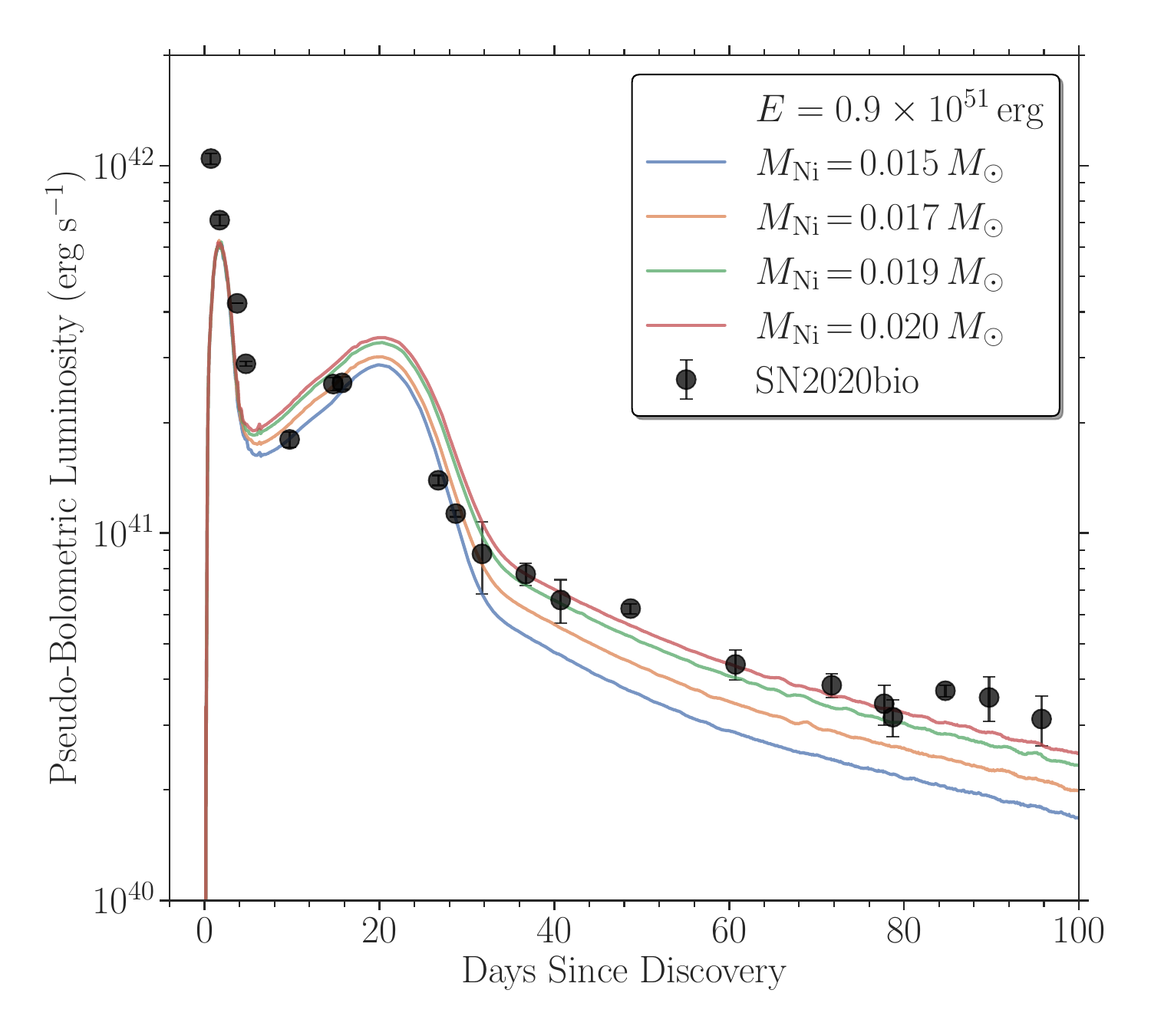}
    \caption{Numerical \texttt{MESA} and \texttt{STELLA} model light curves of SN\,2020bio for varying $M_{\text{Ni}}$. Both the secondary light-curve peak and late-time light-curve slope are best reproduced with $\approx$ 0.02 $M_\odot$ of $^{56}$Ni synthesized in the explosion.}
    \label{fig:numericalmodels}
\end{figure}

SCE dominates the total luminosity only for several days after explosion. The rest of the light curve is powered by the radioactive decay of $^{56}$Ni and its children isotopes. Using our multiband coverage of SN\,2020bio for $\approx$ 160 days after explosion, we construct a pseudo-bolometric light curve to fit for the amount of $^{56}$Ni produced in the explosion. For epochs with observations in more than three filters, we extrapolate the SED out to the blue and red edges of the \textit{U}- and \textit{i}-band filters, respectively, using a univariate spline. We choose to extrapolate the (extinction-corrected) photometry rather than fit a blackbody SED, because the spectra are not representative of a blackbody throughout the object's evolution.

To infer the properties of the pre-explosion progenitor as well as the explosion itself, we compare numerical \texttt{MESA} \citep{Paxton2011,Paxton2013,Paxton2015,Paxton2018,Paxton2019} and \texttt{STELLA} \citep{Blinnikov1998,Blinnikov2000,Blinnikov2006} model explosions to our pseudo-bolometric light curve. As no model from the grid of \citet{Hiramatsu2021} reproduces the weak secondary peak of SN\,2020bio, we expand their model grid with higher wind efficiencies ($\eta$ = 5.0—5.2). For the best-fit models presented here, we begin with a \texttt{MESA} single-star progenitor with zero-age main-sequence (ZAMS) mass $M_{\text{ZAMS}}$ = 15 $M_\odot$ at solar metallicity and evolve it to a final mass of 4.8 $M_\odot$ with a wind efficiency $\eta$ = 5.2 and no rotation. At explosion, the progenitor has a H-rich envelope radius of 280 $R_\odot$ and mass of 0.10 $M_\odot$, in agreement with values we find from our best-fit H-rich SCE models. The explosion energy and ejecta mass are fixed at 0.9 $\times$ 10$^{51}$ erg and 2.9 $M_\odot$, respectively, and the mass of $^{56}$Ni (M$_{\text{Ni}}$) is varied between 0.015 and 0.020 $M_\odot$. These explosion models are then run through \texttt{STELLA} using 600 spatial zones and 100 frequency bins in order to reproduce the bolometric luminosity evolution. For more information, see \citet{Hiramatsu2021}. The \texttt{MESA} model files are available on Zenodo\footnote{\dataset[10.5281/zenodo.7927189]{\doi{10.5281/zenodo.7927189}}}.

The resulting model light curves are shown in Figure \ref{fig:numericalmodels}, compared with the pseudo-bolometric light curve of SN\,2020bio. We find decent qualitative agreement between the numerical models and the observed light-curve evolution, in particular at later times. The secondary light-curve peak and late-time light-curve slope are well reproduced by an explosion that synthesizes $\approx$\,0.02 $M_\odot$ of $^{56}$Ni. The secondary light-curve peak may be overproduced, but the exact peak luminosity and time of peak are uncertain, given the gap in our observational coverage. 

\begin{figure}
    \centering
    \includegraphics[width=0.48\textwidth]{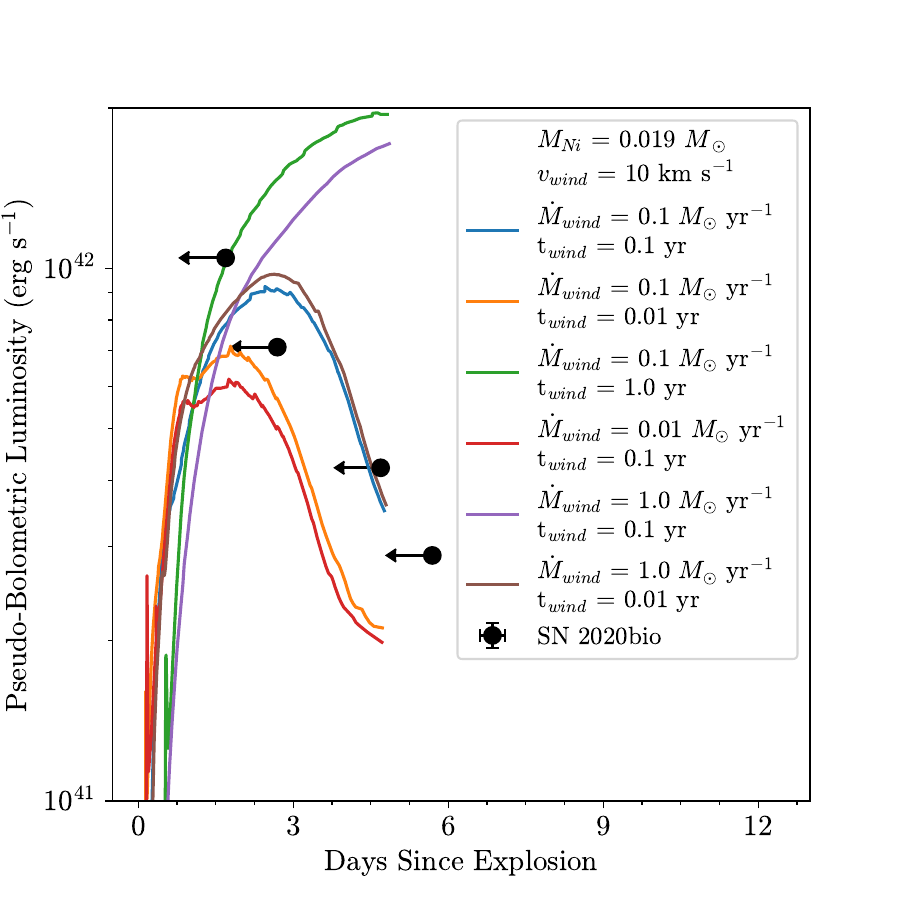}
    \caption{Numerical \texttt{MESA} and \texttt{STELLA} circumstellar interaction-powered model light curves of SN\,2020bio at early times. Different color curves correspond to models with varying mass-loss rates and timescales. Leftward-facing arrows show the range of possible explosion epochs inferred from our SCE fits. The early-time emission excess is best reproduced with 0.001-0.01 $M_\odot$ of CSM.}
    \label{fig:csmmodels}
\end{figure}

Interestingly, however, the peak luminosity of the SCE is not reproduced by these models. It may be that the treatment of the SN shock and the subsequent cooling of the outer envelope is too complex to fully simulate within these models. On the other hand, it is possible that an additional powering mechanism contributes to the early-time evolution. To test this, we explore how the addition of different mass-loss rates and timescales to the models affects the early-time light curve through short-lived circumstellar interaction. To the best-fit \texttt{MESA} model, we attach a wind density profile $\rho_{\text{CSM}}(r) = \dot{M}_{\text{wind}} / 4\pi r^2 v_{\text{wind}}$, where $v_{\text{wind}}$ = 10 km s$^{-1}$, with 100 additional spatial zones. These CSM models are shown in Figure \ref{fig:csmmodels}. We find that the best-fit models have a confined CSM with masses of 1 $\times$ 10$^{-3}$ -- 1 $\times$ 10$^{-2}$ $M_\odot$ lost by the progenitor within the last several months before explosion. This hints that circumstellar interaction may contribute to the rapidly fading early-time emission of SN\,2020bio and possibly other SNe IIb. If this is the case, then the information estimated through SCE model fits may not be truly representative of the true nature of their progenitors.

The values inferred from this numerical modeling, in particular the $^{56}$Ni mass, are on the low end of the distribution of values estimated for other well-studied SNe IIb. SNe IIb with double-peaked light curves typically display secondary radioactive decay-powered peaks equally or more luminous than the peak of the SCE, implying a greater amount of $^{56}$Ni synthesized. Studies using samples of these objects have found average $^{56}$Ni masses of $\approx$ 0.10 -- 0.15 $M_\odot$ and average ejecta masses of $\approx$ 2.2 -- 4.5 $M_\odot$ \citep{Lyman2016,Prentice2016,Taddia2018}, in better agreement with ejecta parameters of other stripped-envelope and H-rich core-collapse SNe. However, rare cases of underluminous SNe IIb with low inferred $M_{\text{Ni}}$ have been discovered \citep[e.g.,][]{Nakaoka2019,Maeda2023}. These objects have light curves that appear transitional between standard SNe II-P and SNe IIb, which differ from the observed photometric evolution of SN\,2020bio. 

On the other hand, in the case of SN\,2018ivc, both a low $^{56}$Ni mass ($M_{\text{Ni}} \leq 0.015 M_\odot$) and progenitor mass ($M_{\text{ZAMS}} \lesssim 12 M_\odot$) are inferred \citep{Maeda2023}. It is possible that other SNe IIb with little synthesized $^{56}$Ni may be undercounted due to their rapidly fading or underluminous light curves. \citet{Maeda2023} also concluded that the light curve of SN\,2018ivc was powered at least in part by circumstellar interaction. Sustained circumstellar interaction has been inferred for other SNe IIb, either through late-time spectral features \citep{Maeda2015,Fremling2019} or through X-ray and radio observations \citep{Fransson1996}. It may be that the mechanism that produced the confined CSM inferred from our numerical models of SN\,2020bio, and possibly that seen in the case of SN\,2018ivc, points to more extreme mass loss than found in other SNe IIb.

\subsection{Comparison to Nebula Spectra Models}\label{subsec:progenitormodels}

\begin{figure*}[t!]
    \centering
    \includegraphics[width=0.8\textwidth]{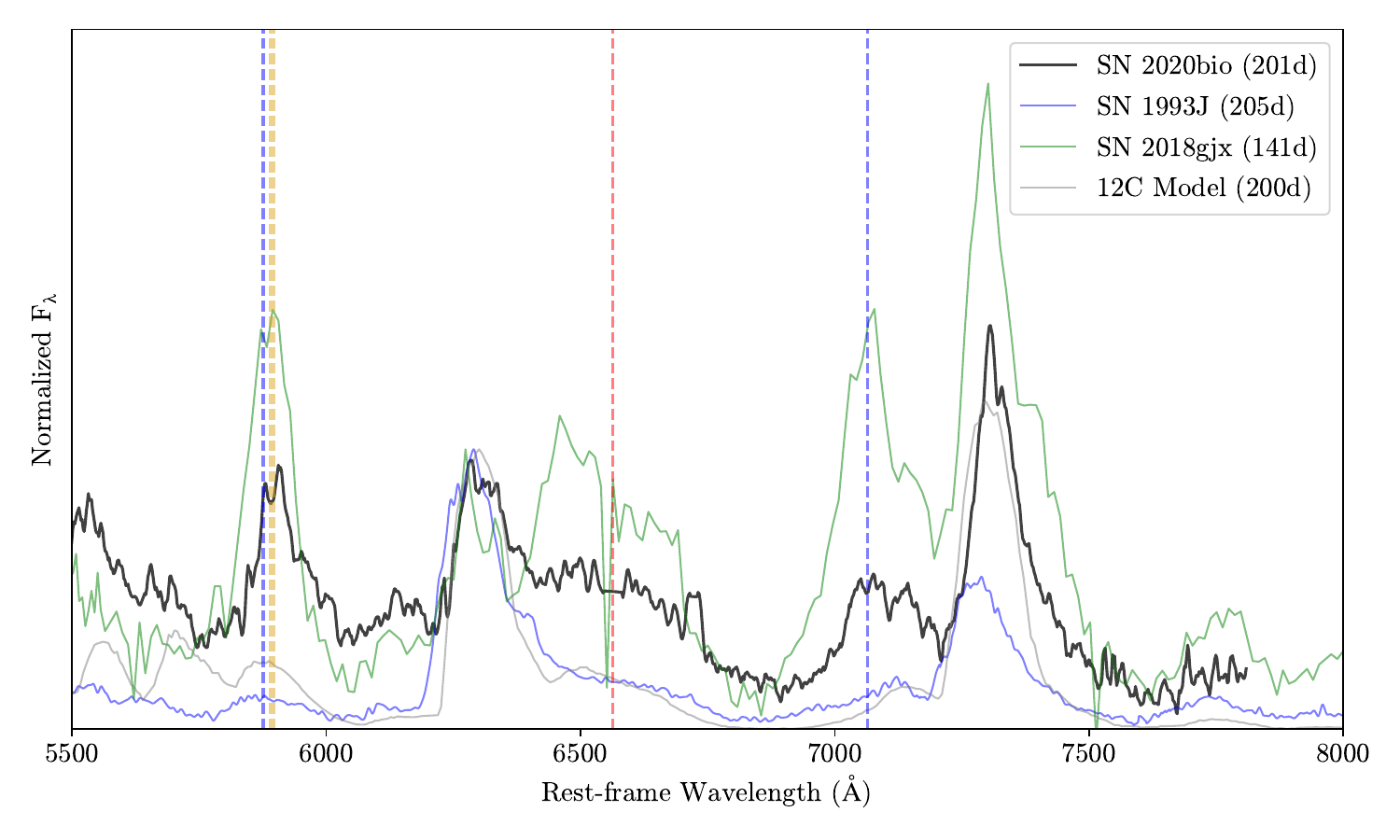}
    \caption{The nebular spectrum of SN\,2020bio (black) compared to that of SN\,1993J (blue), the 12C model of \citet[][gray]{Jerkstrand2015}, and the transitional Type IIb/Ibn SN\,2018gjx \citep{Prentice2020}. Phases are given in the legend. Fluxes have been normalized to the [\ion{O}{1}] emission feature, and galaxy emission lines have been masked for clarity. Notable features in the spectrum of SN\,2020bio have been marked with dashed lines\textemdash H$\alpha$ (red), \ion{He}{1} $\lambda$ 5876, $\lambda$ 7065 (blue), and Na ID (gold).}
    \label{fig:nebspec}
\end{figure*}

A trend between an increasing amount of synthesized O and increasing core-collapse SN progenitor mass has been extensively studied \citep[e.g.,][]{Woosley2007}. \citet{Jerkstrand2015} use this relationship to calibrate the [\ion{O}{1}] $\lambda \lambda$ 6300,6364 luminosity, normalized by the radioactive decay luminosity at the same phase, with numerical models of SNe IIb progenitors \citep[see Eq. 1 of ][]{Jerkstrand2015}. The authors consider models with zero-age main-sequence masses between 12 $M_\odot$ and 17 $M_\odot$. Comparing the observed normalized [\ion{O}{1}] luminosity for a handful of SNe IIb, such as SN\,1993J, SN\,2008ax, and SN\,2011dh, to these models allows for a direct estimate of their progenitor masses\textemdash all of which fall in the range of masses modeled.

Here, we reproduce this analysis using a nebular spectrum of SN\,2020bio, obtained 201 days after the estimated explosion, shown in Figure \ref{fig:nebspec}. We estimate the luminosity from the [\ion{O}{1}] $\lambda \lambda$ 6300,6364 emission doublet in the same way as \citet{Jerkstrand2015}\textemdash assuming the width of the feature to be 5000 km s$^{-1}$, we estimate the continuum by finding the minimum flux redward and blueward of this width and calculate the luminosity within the continuum-subtracted feature. We normalize this luminosity using the luminosity of $^{56}$Ni decay, assuming the best-fit $M_{\text{Ni}}$ = 0.019 $M_\odot$ from Section \ref{subsec:nilc}. 

The normalized luminosity at 201 days is $L_{\text{norm}}$($t$=201)\,=\,9$\times$10$^{-4}$ $\pm$ 2$\times$10$^{-5}$. This value is lower than any of the numerical models analyzed by \citet{Jerkstrand2015}, implying a progenitor mass $\leq$ 12 $M_\odot$. A low progenitor mass for SN\,2020bio can also be inferred from the ratio of the [\ion{Ca}{2}] $\lambda \lambda$ 7311, 7324 to [\ion{O}{1}] $\lambda \lambda$ 6300, 6364 fluxes. A higher ratio implies a lower-mass progenitor, with SNe IIb from literature having values $\lesssim$ 1 throughout their nebular phases \citep[e.g.,][]{Fang2019,Terreran2019,Hiramatsu2021}. Using the same procedure as above, we estimate a [\ion{Ca}{2}] to [\ion{O}{1}] ratio of 1.34 $\pm$ 0.03\textemdash again pointing to a low-mass progenitor star.

Nebular spectroscopy has also been used to infer the presence of late-time circumstellar interaction in several SNe IIb, including SN\,1993J \citep{Fransson1996}, SN\,2013df \citep{Maeda2015}, and ZTF18aalrxas \citep{Fremling2019}. In these objects, interaction with H-rich material lost by the progenitor star was inferred through the presence of a boxy H$\alpha$ profile that strengthened with time. While the origin of this feature at times $\lesssim$ 300 days after explosion is debated \citep[e.g.,][]{Fang2018}, \citet{Fremling2019} show that this feature is visible as early as $\approx$ 180 days after explosion. To search for signatures of interaction, in Figure \ref{fig:nebspec} we compare the nebular spectrum of SN\,2020bio with that of SN\,1993J from a similar phase, the 12C model from \citet{Jerkstrand2015}, and the interacting Type IIb/Ibn SN\,2018gjx \citep{Prentice2020}. The host galaxy emission lines have been masked in the spectrum of SN\,2020bio and all spectra have been normalized to the strength of the [\ion{O}{1}] emission feature. The relative strength of the [\ion{O}{1}] and [\ion{Ca}{2}] features of SN\,2020bio is reproduced well by this model, again supporting a $\approx$ 12 $M_\odot$ zero-age main sequence progenitor.

At this phase, SN\,2020bio exhibits several unusual features. An excess redward of the [\ion{O}{1}] $\lambda$$\lambda$ 6300,6364 feature relative to the 12C model can be attributed to H$\alpha$ powered by circumstellar interaction \citep{Maeda2015,Fremling2019}. While this feature may be due to [\ion{N}{2}] \citep{Jerkstrand2015}, \citet{Fremling2019} found that ZTF18aalrxas still showed an excess even after subtracting off numerical models of SN IIb nebular spectra at this phase. More intriguing are two additional features centered around the \ion{He}{1} $\lambda$ 5876 and \ion{He}{1} $\lambda$ 7065 lines. These features are not found in the 12C model spectrum; however, they are of comparable width (albeit weaker in intensity) to the \ion{He}{1} lines found in the nebular spectrum of SN\,2018gjx. \citet{Prentice2020} claim that these lines are due to persistent interaction with He-rich circumstellar material that is revealed after the SN photosphere has receded sufficiently far into the ejecta. It is interesting to note that SN\,2020bio and SN\,2018gjx share several characteristics, including: H$\alpha$ P Cygni features that are weaker than that of SN\,1993J roughly 20 days after explosion; strong shock-cooling emission and a weak secondary light-curve peak, with similar amounts of $^{56}$Ni produced; and weak [\ion{O}{1}] compared to [\ion{Ca}{2}] in their nebular spectra.

Based on its low synthesized $^{56}$Ni mass and nebular spectral features, we conclude that SN\,2020bio was likely the core collapse of a star with a lower mass than the progenitors of most other SNe IIb surrounded by an unusually massive CSM at the time of explosion.

\section{Discussion and Conclusions}\label{sec:discussion}

We have presented rapid multiband photometric and spectroscopic observations of SN\,2020bio, a Type IIb SN with luminous and rapidly evolving SCE, beginning $\leq$ 1 day after explosion. Compared with other well-observed SNe IIb, SN\,2020bio has the bluest colors at early times as well as weak H spectral features throughout its evolution. Fitting analytical models of SCE to the early-time light curve gives progenitor radii on the order of 100 $R_\odot$ -- 500 $R_\odot$ and envelope masses of 0.01 $M_\odot$ -- 0.5 $M_\odot$ for our best-fit models, which are slightly greater than values derived for other SNe IIb progenitors using the same methods \citep[e.g., SN\,2016gkg;][]{Arcavi2017}. The weak secondary peak powered by radioactive decay is evidence of relatively little $^{56}$Ni synthesized, $M_{\text{Ni}}$ $\approx$ 0.02 $M_\odot$, which is in tension with average $M_{\text{Ni}}$ estimates from samples of other SNe IIb. Numerical modeling of the progenitor explosion within confined circumstellar material is consistent with the observed light curve, showing that circumstellar interaction is likely needed to reproduce the complete pseudo-bolometric light curve. Finally, comparing the nebular spectra to numerical models implies a progenitor 
ZAMS mass $\leq$ 12 $M_\odot$ and reveals signatures of interaction over 200 days after explosion. 

It is difficult to explain all these peculiar features of SN\,2020bio in one consistent model. The combination of its blue colors, spectral features, and our numerical modeling points to interaction with confined CSM that was stripped from the progenitor's outer envelope during the months prior to explosion. The best-fit progenitor parameters, in particular the large envelope radius and low envelope mass, may suggest an inflated progenitor undergoing enhanced mass-loss immediately before exploding. However, the very low $^{56}$Ni and ejecta masses inferred from the later-time light curve, as well as the nebular spectroscopy, point to a lower-mass progenitor. It is possible that SN\,2020bio was the collapse of an unusually low-mass core within a dense CSM produced from its lost H layers. Such extensive mass-loss may require interaction with a binary companion, as inferred for other SNe IIb \citep[e.g.,][]{Maund2004,Benvenuto2013,Prentice2020}, or may be more consistent with wave-driven outbursts caused by core neon burning in stars with M$_{\text{ZAMS}} \approx$ 11 $M_\odot$ \citep{Wu2021,Wu2022}. Interaction between the SN ejecta and this CSM explains the blue colors and obscured ejecta features at early times, while the small $^{56}$Ni mass and nebular spectrum indicate a low zero-age main-sequence mass. This interaction can lead to an overestimated progenitor radius\textemdash if the CSM was near enough to the progenitor, we may have observed the shock cooling of this extended CSM instead of the outer envelope of the progenitor. In the future, more detailed models and multiwavelength observations, in particular in the radio and X-rays, will be needed in order to infer SNe IIb progenitor mass-loss rates and CSM masses.

Given the weak H spectral features when compared to spectra of other SNe IIb, SN\,2020bio may be an intermediary object between the Type IIb and Ib subclasses, representing a progenitor that was recently stripped almost entirely of its H-rich envelope. Transitional objects between SNe IIb and SNe Ib have been observed \citep{Prentice2017} and can be explained by different amounts of H remaining in the outer envelope at the time of explosion. More difficult to explain are the small $^{56}$Ni and ejecta masses, which are lower than those measured for both SNe IIb and SNe Ib \citep[e.g.,][]{Taddia2018}. Some objects that exist in the literature with both low ejecta and $^{56}$Ni masses and observed SCE are peculiar SNe Ib as well as Ca-rich transients. However, it is difficult to reconcile the photospheric-phase spectra of SN\,2020bio, which are most similar to those of other SNe IIb, with the spectra of these objects, which are often used to argue for a degenerate or ultra-stripped progenitor \citep{Yao2020,WJG2022}. Instead, it is more likely that SN\,2020bio had a massive star progenitor more similar to the progenitors of other SNe IIb based on their similar photospheric-phase spectral features. 

Perhaps more interesting are the similarities between SN\,2020bio and SN\,2018gjx. SN\,2018gjx is a peculiar object; \citet{Prentice2020} found that it showed evidence for CSM shock cooling at early times before displaying typical SN IIb spectra. Around 30 days after explosion, the spectra began showing strong emission features of \ion{He}{1} that the authors argued were powered by interaction with He-rich material, revealed by the receding photosphere, that persisted into the nebular phase. Like SN\,2020bio, it had a weak secondary light-curve peak that implied a synthesized $^{56}$Ni mass of 0.021 $M_\odot$. A scenario that explains all these observations is an SN IIb explosion with an asymmetric ring or torus of CSM, viewed equatorially. The fact that the interaction signatures are not as prevalent in the spectra of SN\,2020bio, both within the first few days of explosion and during the nebular phase, may suggest a viewing angle between equatorial and polar\textemdash if this is the case, more of the SN photospheric and nebular features would be visible, rather than obscured by the ongoing interaction. The CSM masses inferred from model fits for both objects are roughly the same ($\approx$ 0.01 $M_\odot$). This may reveal that the progenitor of SN\,2020bio underwent extensive mass loss, perhaps losing almost all of its H-rich material before exploding. Even in this extreme case, however, the models used throughout this analysis \citep[e.g.,][]{Piro2015,Sapir2017,Piro2021} have been applied to both H-rich and H-poor objects. Therefore, we are confident our conclusions remain valid, regardless of the exact amount of H remaining in the progenitor envelope.

This study contributes to the overall diversity in the progenitors of SNe IIb. More systematic studies of SNe with observed SCE will be needed in order to search for similarities and differences in their progenitor systems. In particular, this work shows the importance of rapid, multiwavelength follow-up of these objects. It is particularly important to better understand the number of SNe IIb with weak secondary light-curve peaks, such as SN\,2020bio. These objects may have later-time ($\geq$ 5 days) luminosity below the detection threshold of current all-sky surveys as well as rapid early-time emission that evolves too quickly to be extensively followed. Therefore, we may be undercounting the rates of core-collapse, stripped-envelope SNe with low $^{56}$Ni and ejecta masses. A better understanding of their progenitors will be important for exploring the low-mass end of core-collapse SNe.

\begin{acknowledgments}
This work made use of data from the Las Cumbres Observatory network. The LCO group is supported by AST-1911151 and AST-1911225 and NASA Swift grant 80NSSC19k1639. I.A. is a CIFAR Azrieli Global Scholar in the Gravity and the Extreme Universe Program and acknowledges support from that program, from the European Research Council (ERC) under the European Union's Horizon 2020 research and innovation program (grant agreement No. 852097), from the Israel Science Foundation (grant No. 2752/19), from the United States - Israel Binational Science Foundation (BSF), and from the Israeli Council for Higher Education Alon Fellowship. 

This work made use of the data products generated by the NYU SN group, and 
released under DOI:10.5281/zenodo.58766, 
available at \url{https://github.com/nyusngroup/SESNtemple/}.

This work has made use of data from the Asteroid Terrestrial-impact Last Alert System (ATLAS) project. The Asteroid Terrestrial-impact Last Alert System (ATLAS) project is primarily funded to search for near-Earth asteroids through NASA grants NN12AR55G, 80NSSC18K0284, and 80NSSC18K1575; byproducts of the NEO search include images and catalogs from the survey area. This work was partially funded by Kepler/K2 grant J1944/80NSSC19K0112 and HST GO-15889, and STFC grants ST/T000198/1 and ST/S006109/1. The ATLAS science products have been made possible through the contributions of the University of Hawaii Institute for Astronomy, Queen’s University Belfast, the Space Telescope Science Institute, the South African Astronomical Observatory, and The Millennium Institute of Astrophysics (MAS), Chile.
\end{acknowledgments}

\software{\texttt{Astropy} \citep{Astropy2018}, \texttt{emcee} \citep{ForemanMackey2013}, \texttt{lcogtsnpipe} \citep{Valenti2016}, Matplotlib \citep{Hunter2007}, \texttt{MESA} \citep{Paxton2011,Paxton2013,Paxton2015,Paxton2018,Paxton2019}, Numpy \citep{Harris2020}, \texttt{STELLA} \citep{Blinnikov1998,Blinnikov2000,Blinnikov2006}}

\appendix

\restartappendixnumbering

\section{Corner Plots}\label{sec:corners}

\begin{figure}[b!]
    \centering
    \includegraphics[width=0.9\textwidth]{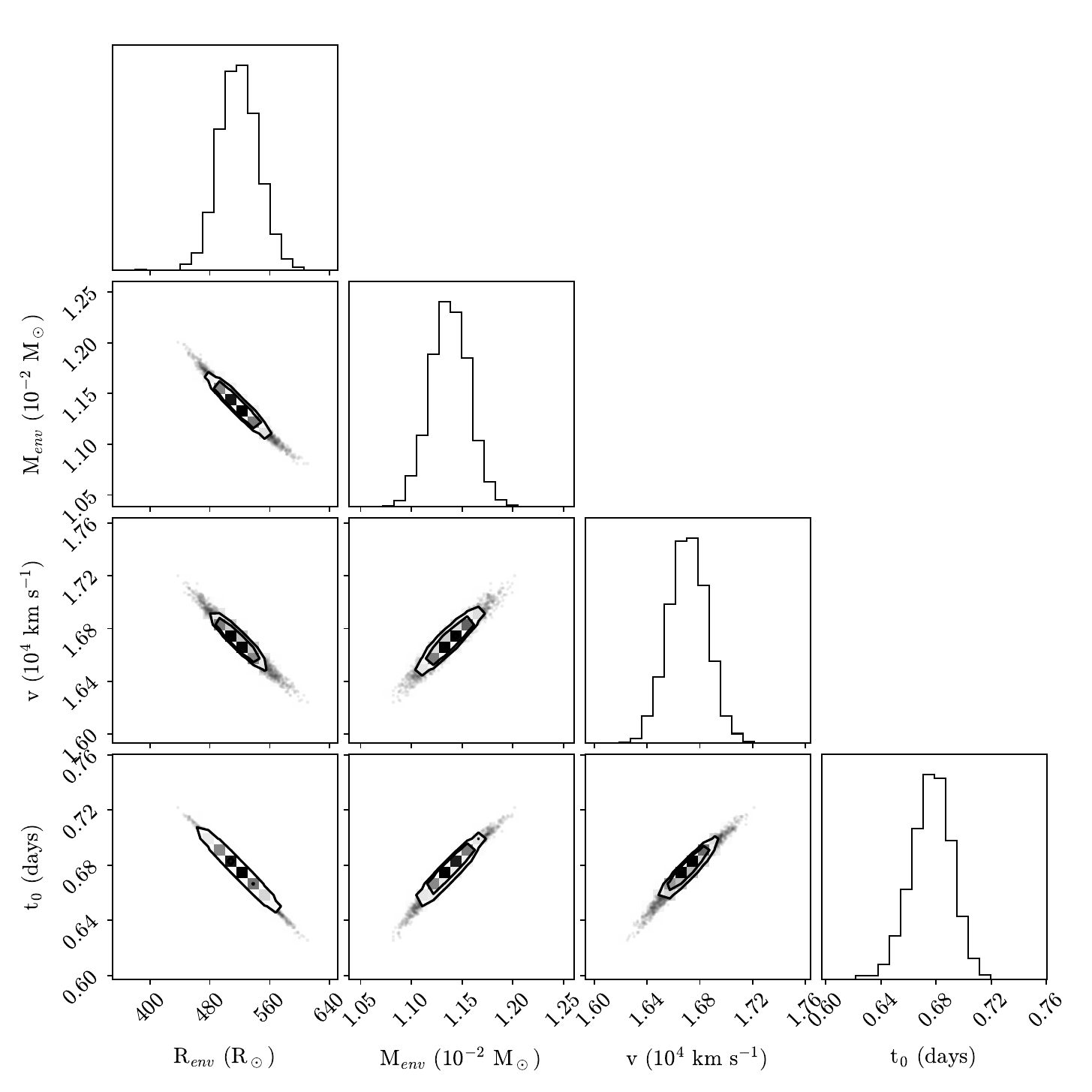}
    \caption{Corner plots showing the fitted parameter distributions for the P15 model. }
    \label{fig:p15corner}
\end{figure}

\begin{figure}
    \centering
    \includegraphics[width=0.9\textwidth]{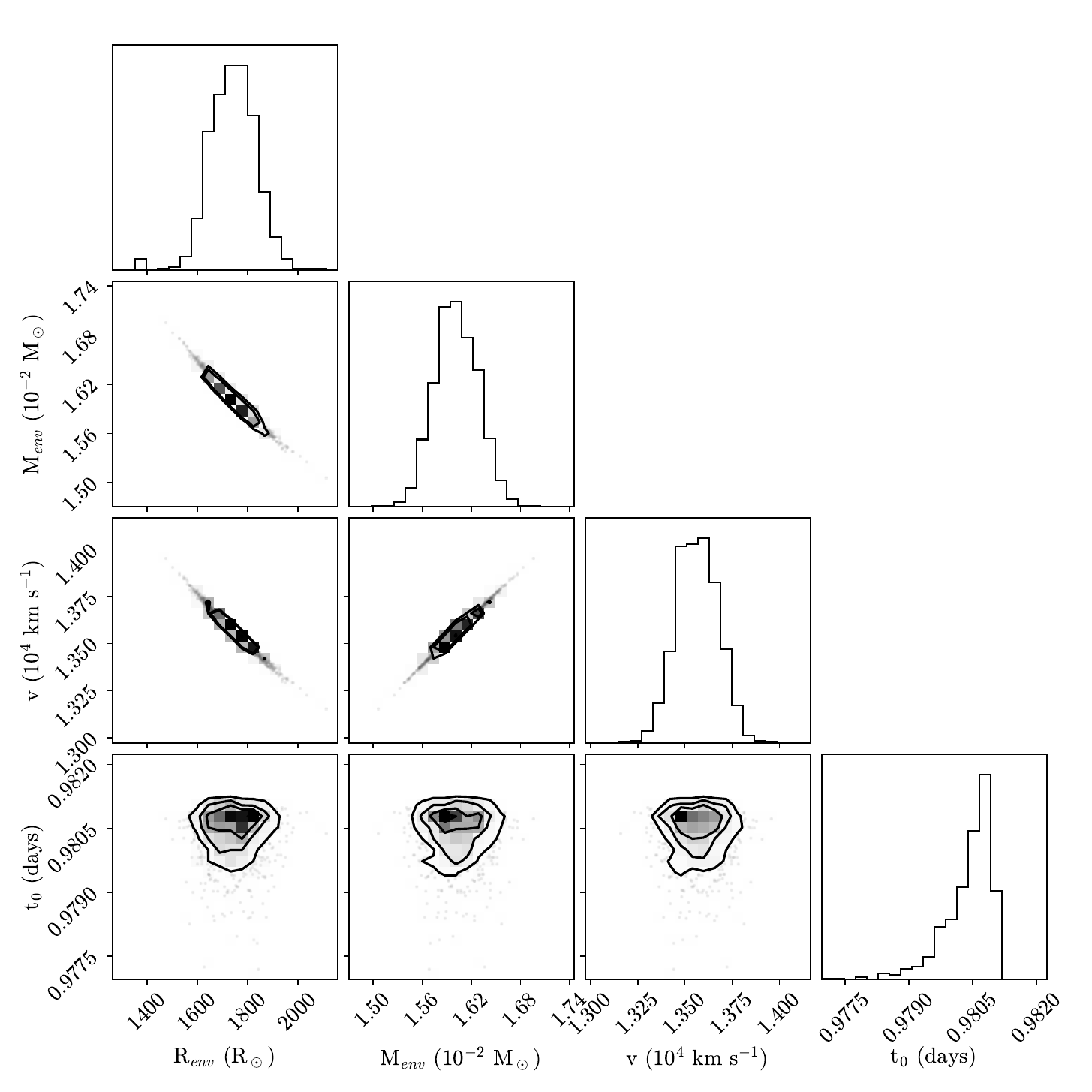}
    \caption{Same as Figure \ref{fig:p15corner}, but for the P21 model.}
    \label{fig:p21corner}
\end{figure}

\begin{figure}
    \centering
    \includegraphics[width=0.9\textwidth]{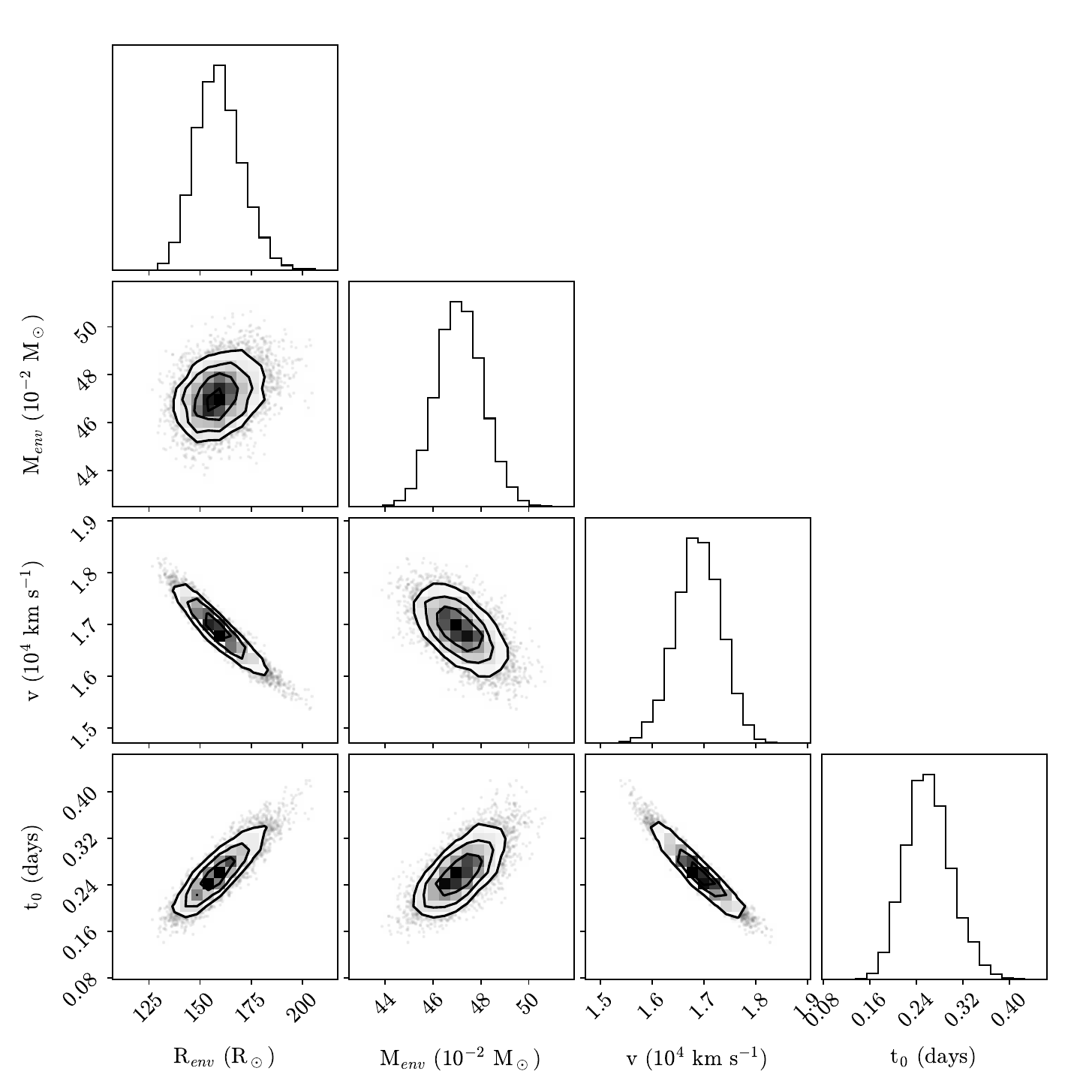}
    \caption{Same as Figure \ref{fig:p15corner}, but for the SW17 ($n$=3/2) model.}
    \label{fig:sw17corner}
\end{figure}

\begin{figure}
    \centering
    \includegraphics[width=0.9\textwidth]{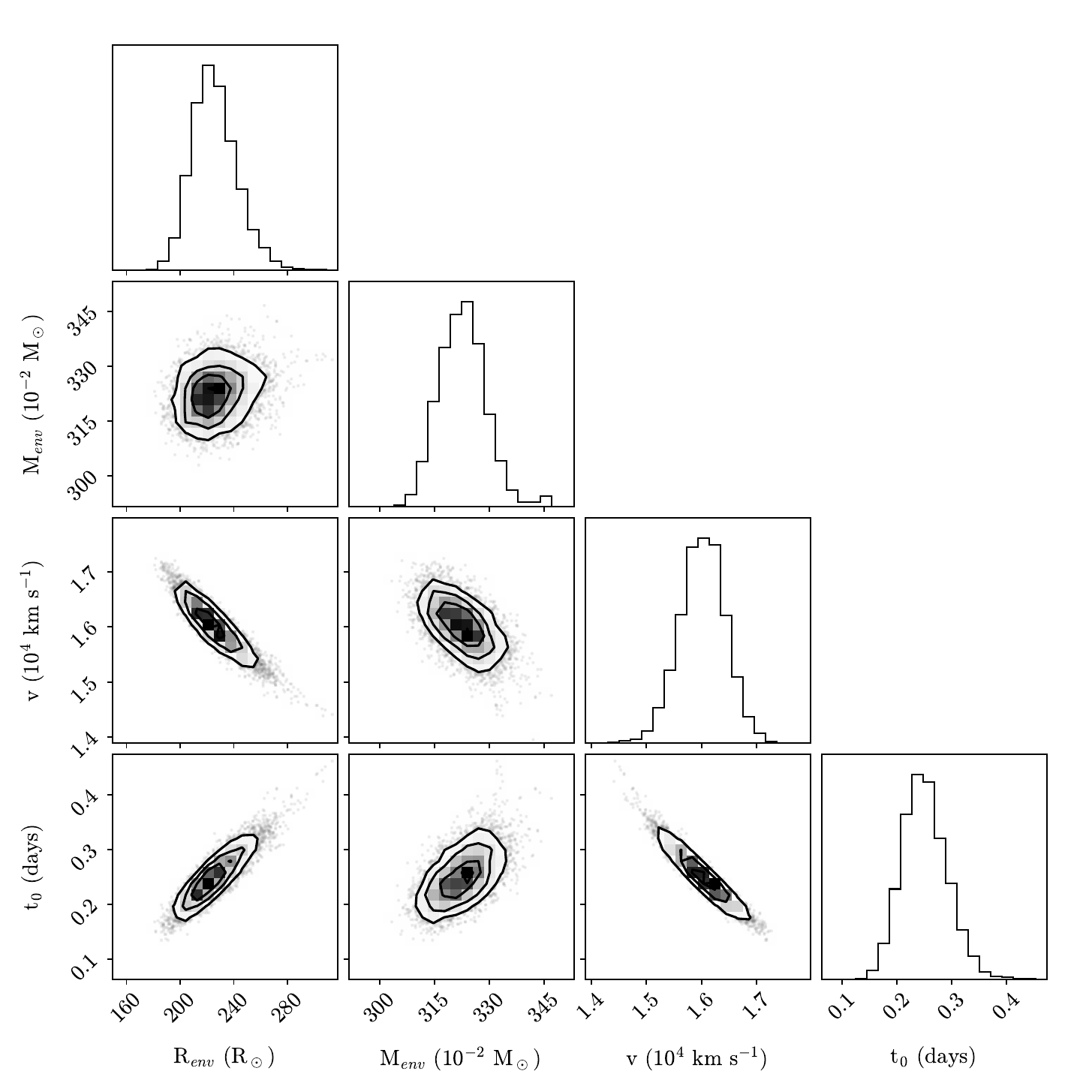}
    \caption{Same as Figure \ref{fig:p15corner}, but for the SW17 ($n$=3) model.}
    \label{fig:sw17_bsg_corner}
\end{figure}

\begin{figure*}[t!]
    \centering
    \includegraphics[width=0.49\textwidth]{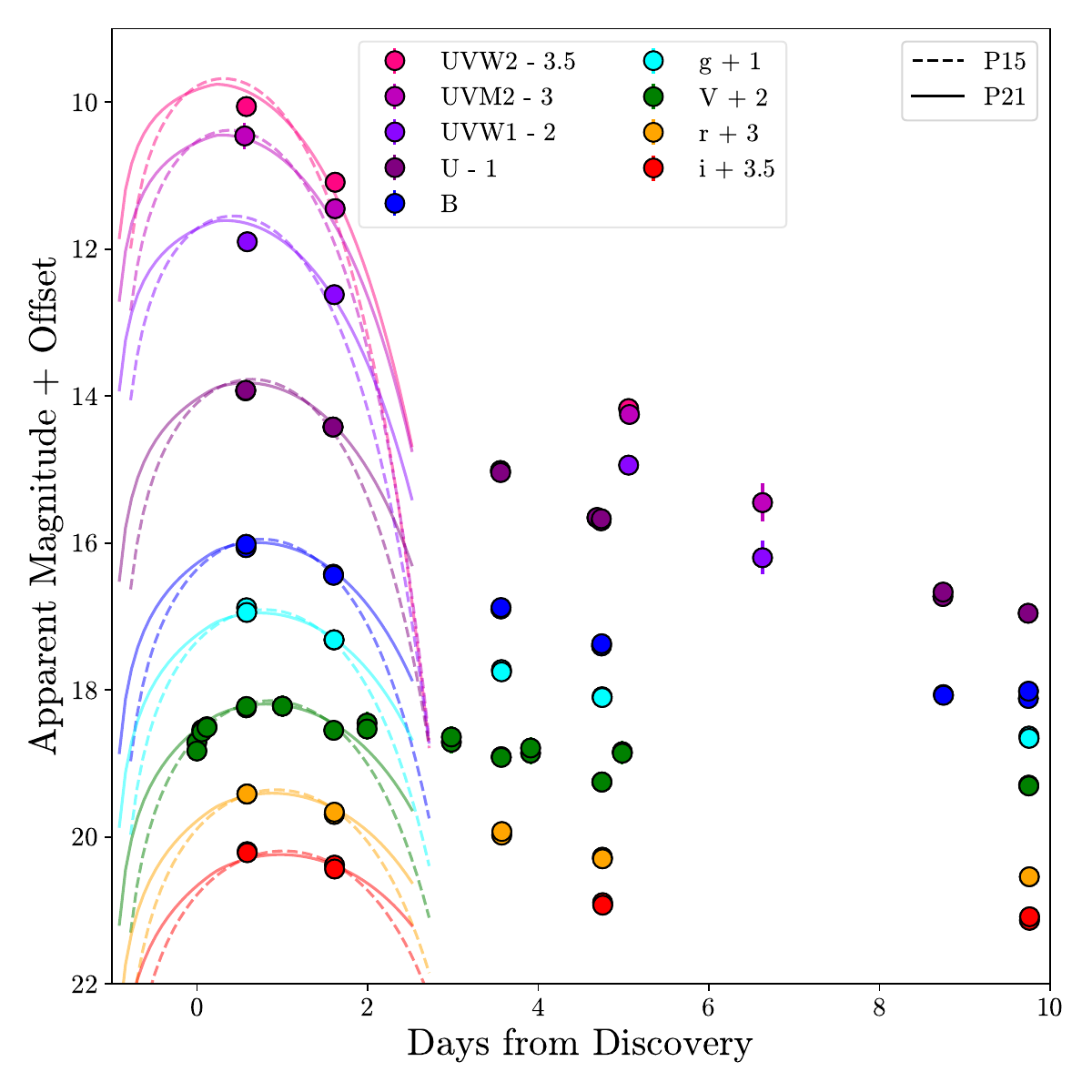}
    \includegraphics[width=0.49\textwidth]{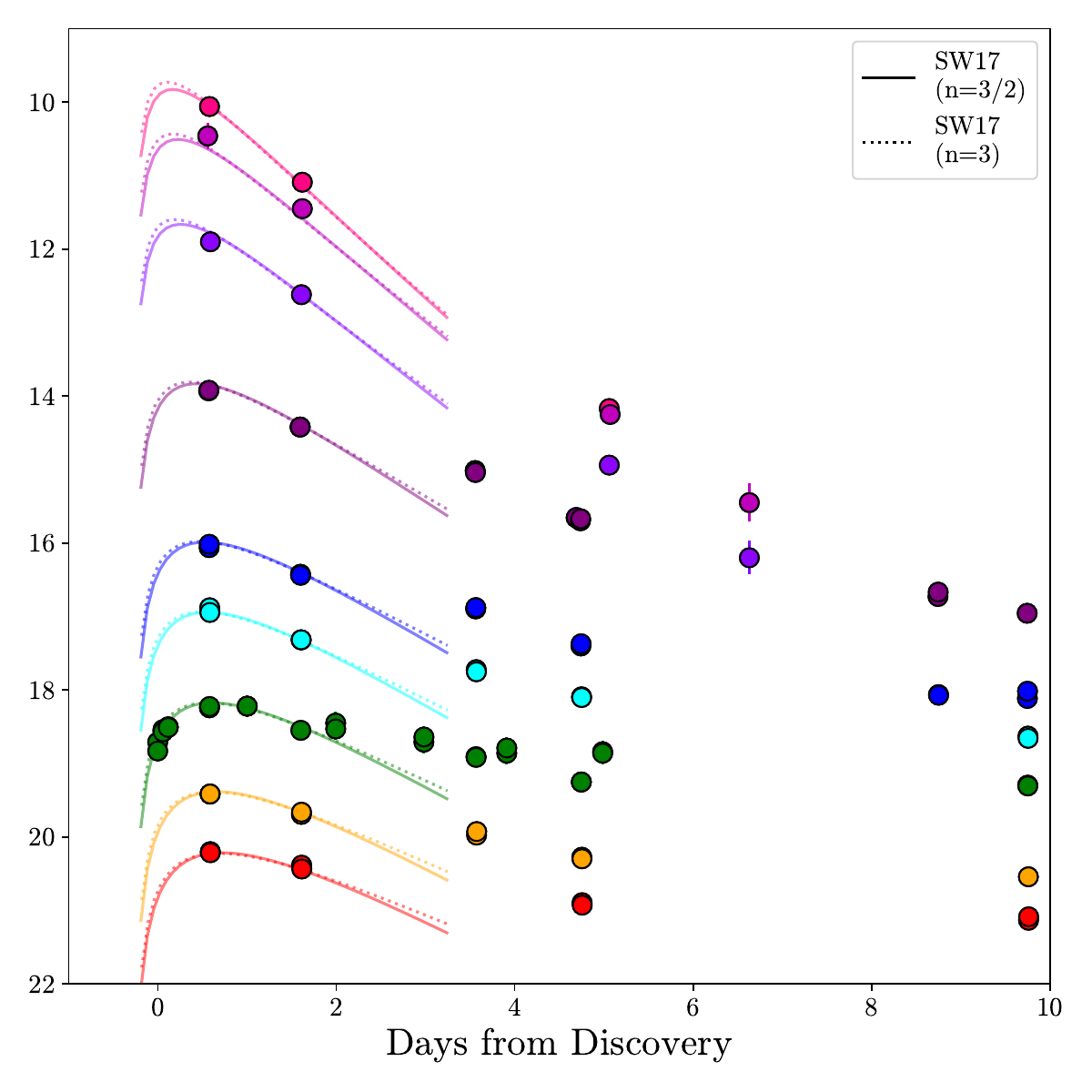}
    \caption{Same as Figure \ref{fig:scefits} but assuming a H-poor envelope composition ($\kappa$ = 0.20 cm$^2$ g$^{-1}$). The fits are of comparable quality to those assuming a H-rich composition, but the best-fit parameters are quantitatively different (Table \ref{tab:hpoormodelparams}).}
    \label{fig:hpoorfits}
\end{figure*}

In Figures \ref{fig:p15corner}, \ref{fig:p21corner}, \ref{fig:sw17corner}, and \ref{fig:sw17_bsg_corner} we present distributions of the fitted parameters of the models detailed in Section \ref{subsec:models}. In Figure \ref{fig:hpoorfits} we present fits assuming a H-poor envelope composition, appropriate for a stripped-envelope SN, with best-fit parameter values given in Table \ref{tab:hpoormodelparams}.

\begin{deluxetable*}{lccccc}[b!]
\tablecaption{SCE H-poor Model Parameters\label{tab:hpoormodelparams}}
\tablehead{
\colhead{Model} & \colhead{$R_{\text{env}}$ ($R_\odot$)} & \colhead{$M_{\text{env}}$ (10$^{-2}$ $M_\odot$)} & \colhead{$v^a$ (10$^{4}$ km s$^{-1}$)} & \colhead{$t_0$ (days)} & \colhead{$\chi^2$ / d.o.f.}}
\startdata
P15 & 309$_{-17}^{+18}$ & 1.94$_{-0.04}^{+0.03}$ & 1.67$_{-0.01}^{+0.02}$ & 0.78$_{-0.01}^{+0.01}$ & 21.6\\
P21 & 1010$_{-81}^{+59}$ & 2.74$_{-0.06}^{+0.07}$ & 1.36$_{-0.01}^{+0.02}$ & 0.98$_{-0.01}^{+0.01}$ & 21.0\\
SW17 ($n$=3/2) & 92$_{-8.4}^{+7.5}$ & 80.9$_{-1.70}^{+2.00}$ & 1.70$_{-0.06}^{+0.05}$ & 0.26$_{-0.04}^{+0.04}$ & 8.7\\
SW17 ($n$=3) & 140$_{-9.7}^{+11}$ & 857$_{-16.1}^{+16.6}$ & 2.50$_{-0.07}^{+0.07}$ & 0.25$_{-0.04}^{+0.05}$ & 8.7\\
\enddata
\tablenotetext{a}{The characteristic velocity for P15 and P21 and the shock velocity for SW17.}
\end{deluxetable*}

\end{document}